\documentclass[epj,nopacs,final]{svjour2}
\usepackage{epsfig} 
\usepackage{longtable} 
\usepackage{hyperref}
\usepackage{bm}

\usepackage{latexsym}
\usepackage{graphics}
\usepackage{multirow}
\usepackage{amssymb}

\newcommand{\PT}{\mbox{$P_T$}}

\newcommand{\ptmiss}{\mbox{$P_T^{miss}$}}

\newcommand{\Gcs}{${\rm{GeV/}c^2}$}
\newcommand{\Tcs}{${\rm{TeV/}c^2}$}
\newcommand{\Gc}{${\rm{GeV/}c}$}
\newcommand{\hwws}{\mbox{$H \rightarrow  W W^{(*)}$}}
\newcommand{\hwwsll}{\mbox{$H \rightarrow  W W^{(*)} \rightarrow \ell \nu \ell \nu $}}

\newcommand{\gamgam}{\mbox{$\gamma \gamma$}}
\newcommand{\htautau}{\mbox{$H \rightarrow \tau \tau$}}

\newcommand{\ttbar}{\mbox{$t \overline{t} $}}
\newcommand{\bbbar}{\mbox{$b \overline{b} $}}

\newcommand{\ttjets}{\mbox{$t \overline{t} + jets$}}

\newcommand{\fbs}{\mbox{$\rm{fb}^{-1}$}}

\newcommand{\ewtau}{EW \mbox{$\tau \tau + jets$}}

\newcommand{\qcdww}{QCD \mbox{$W W + jets$}}
\newcommand{\ewww}{EW \mbox{$W W + jets$}}
\newcommand{\wwjets}{\mbox{$W W + jets$}}

\newcommand{\sig}{\mbox{$\sigma$}}

\newcommand{\hwwll}{\mbox{$H \rightarrow W W^{(*)} \rightarrow \ell \nu \ell \nu$}}

\begin{document}

\title{Prospects for the Search for a Standard Model Higgs Boson \\
in ATLAS using Vector Boson Fusion}

\author{S.Asai\inst{8}, G.Azuelos\inst{5} 
C.Buttar\inst{7}, V.Cavasinni\inst{6}, 
D.Costanzo\inst{6}\fnmsep\thanks{Now at LBNL, Berkeley, USA}, K.Cranmer\inst{9},
R.Harper\inst{7},  
K. Jakobs\inst{4}, J.Kanzaki\inst{3}, M.Klute\inst{1}, 
R.Mazini\inst{5}, B.Mellado\inst{9}, 
W.Quayle\inst{9}, 
E.Richter-W\c{a}s\inst{2},
T.Takemoto\inst{3}, I.Vivarelli\inst{6}, 
Sau Lan Wu\inst{9}}

\authorrunning{S.Asai et al.}

\institute{
Physikalisches Institut, Universit\"at Bonn, Germany.\and
CERN, Geneva, Switzerland\thanks{On leave of absence from 
Jagellonian University, Cracow, Poland} and 
Inst. of Nuclear Physics, Cracow, Poland.  \and    
High Energy Research Organisation (KEK), IPNS, Japan. \and
Institut f\"ur Physik, Universit\"at Mainz, Germany. \and
University of Montreal, Canada. \and
INFN and University of Pisa, Italy. \and
Dept. of Physics and Astronomy, University of Sheffield, UK.\and
University of Tokio, Tokio, Japan.\and
University of Wisconsin-Madison, Wisconsin, USA.
}

\date{Received: 24 April 2003 / Published online: 30 July 2003}

\abstract{The potential for the discovery of a Standard Model Higgs boson 
in the mass range $m_H < 2 m_Z$ in the vector boson fusion mode has 
been studied for the ATLAS experiment at the LHC.
The characteristic signatures of additional jets in the 
forward regions of the detector and of low jet activity 
in the central region allow for an efficient background rejection. 
Analyses for the $\hwws$ and 
$H \rightarrow \tau\tau$ decay modes have been performed 
using a realistic simulation of the expected detector performance.
The results obtained demonstrate the large discovery potential in the \hwws\ decay
channel and the sensitivity to Higgs boson decays into $\tau$-pairs in the 
low-mass region around 120 \Gcs.
}

\maketitle

\section{Introduction}

The search for the Higgs boson is one of the primary tasks of the 
experiments at the {\em Large Hadron Collider} (LHC). It has been established 
by many studies \cite{atlas-tdr,cms-tp}
that a Standard Model Higgs boson can be discovered with high significance 
over the full mass range of interest, from the lower limit set by the 
LEP experiments of 114.1~\Gcs\  \cite{LEP-limit} up to about 1~\Tcs.

At the LHC, the production cross-section for a Standard Model Higgs boson 
is dominated by gluon-gluon fusion. The second largest cross-section comes
from the fusion of vector bosons radiated from initial-state quarks. 
The relative contributions of the two processes depend
on the Higgs boson mass. For $m_H < 2 m_Z$,
vector boson fusion amounts in leading order to about 20\% of the 
total production cross-section and becomes more important
with increasing mass. 
However, for this production mode, additional event characteristics can 
be exploited to suppress the large backgrounds. 
The Higgs boson is accompanied by two jets in the 
forward regions of the detector, originating from the 
initial quarks that emit the vector bosons. In addition, central jet 
activity is suppressed due to the lack of color exchange between the  
quarks. This is in contrast to most background processes, where there is
color flow in the $t$-channel. Therefore
jet tagging in the forward region of the detector together and a veto 
of jet activity in the central region are useful tools 
to enhance the signal-to-background ratio. These techniques have been considered already in the 
studies performed for heavy Higgs bosons \cite{atlas-tdr}. 

The observation of the Standard Model Higgs boson at the LHC 
in the vector boson fusion channels in the intermediate mass range 
was first discussed in Refs.~\cite{zeppenfeld} and \cite{zeppenfeld-ww} 
for the 
$H \rightarrow \gamma \gamma $ and 
$\hwws$  decay modes and in Ref.~\cite{zeppenfeld-tau} for the 
$\htautau$ decay mode. 
The latter 
is particularly interesting for a measurement of the Higgs boson coupling to 
fermions in the low-mass region, {\em i.e.,} $m_H < 140$~\Gcs, since beyond 
the $\bbbar$ decay mode, no other direct fermion decay mode is accessible at
the LHC.

In the present study, the analyses for the $WW^{(*)}$ and 
$\tau\tau$ decay modes have been performed
using realistic simulations of the expected performance of the ATLAS
detector at the LHC, including forward jet tagging.
The performance is addressed at low LHC 
luminosity, {\em i.e.,}
${\cal L} =  10^{33}$  $\rm{cm}^{-2} \rm{sec}^{-1}$,
and the discovery potential 
is evaluated for an integrated luminosity up to 30 $\fbs$, which is 
expected to be reached during the first few years of operation.

In Section 2, the cross-sections and the event generation
for the 
signal and for various background processes are discussed. Important 
experimental issues in the search for vector boson fusion processes at 
the LHC are addressed in Section 3. General signal selection criteria are 
presented in Section 4, before the detailed analyses for the 
$\hwws$ and for the $H \rightarrow \tau \tau$ decay modes are considered in 
Sections 5 and 6. The ATLAS Higgs boson discovery potential in the low-mass 
region is summarized in Section~7.

\section{Signal and Backgrounds}
The cross-sections for the vector boson fusion process have 
been calculated 
using the programme VV2H \cite{spira-hqq}. The results are given 
in Table \ref{t:sig_br} as a function of the Higgs boson mass. Although 
next-to-leading order calculations are available
\cite{fusion-nlo} and have been found to increase the cross-section by 
$\sim 10\%$, leading order cross-sections have been used.
The main reason for this approach is the consistency with the background
estimates, for which NLO cross-section calculations are not available 
for all relevant processes. 
Since the NLO correction for the signal is relatively small, 
the significance might be overestimated if the NLO corrections for the
backgrounds are large. 
The products of the cross-sections multiplied by
the branching ratios of the Higgs boson into $WW$ and $\tau \tau$, which 
have been calculated using the programme HDECAY \cite{hdecay}, are also 
included in Table~\ref{t:sig_br}. 

\begin{table*}
\begin{center}
\begin{minipage}{.75\linewidth} 
\footnotesize
\begin{tabular}{l r || c | c | c | c | c | c | c | c }
\hline
\hline
$m_H$ & (\Gcs) & 120 & 130 & 140 & 150 & 160 & 170 & 180& 190 \\
\hline
\hline
$\sigma (qq H)$ & (pb) & 4.36 & 4.04 & 3.72 & 3.46 & 3.22 & 3.06 & 2.82 & 2.64 \\
\hline
$\sigma \cdot BR (H \rightarrow WW^{(*)})$ & (fb) & 
531 & 1127 & 1785 & 2370 & 2955 & 2959 & 2620 & 2054 \\
$\sigma \cdot BR (H \rightarrow \tau \tau)$ & (fb) & 
 304 & 223 & 135 & 64.4 & 11.9 & 2.8 & 1.6 & 1.0 \\
\hline
\hline
\end{tabular}
\footnotesize
\caption{\small \it Total vector boson fusion production cross-sections $\sigma (qqH)$ 
and 
$\sigma \cdot BR (H \rightarrow W W^{(*)})$ and 
$\sigma \cdot BR (H \rightarrow \tau \tau )$ as a function of the Higgs boson mass. 
The cross-sections have been computed using the CTEQ5L structure function
parametrization~\cite{cteq-sf}.
}\label{t:sig_br}
\end{minipage}
\end{center}
\end{table*}

The following decay chains have been considered in the analysis: 
\begin{itemize}
\item $  qq H \rightarrow qq \ W^+ W^- \rightarrow \  qq \ \ell^+ \nu \ \ell^- \nu $ 
\item $  qq H \rightarrow qq \ W^+ W^- \rightarrow \  qq \ \ell^{\pm} \nu \ jet \ jet$ 
\item $  qq H \rightarrow qq \ \tau^+ \tau^-$ \hspace*{0.29cm}
$\rightarrow \  qq \ \ell^+ \nu \nu \ \ell^- \nu \nu $ 
\item $  qq H \rightarrow qq \ \tau^+ \tau^-$ \hspace*{0.29cm} $\rightarrow \  
qq \ \ell^{\pm} \nu \nu \ \ had \ \nu $. 
\end{itemize}

All final states consist of at least one high-$\PT$ lepton, missing 
transverse momentum and two jets in the forward regions of the detector. 
The following background processes are common to all channels considered: 

\begin{itemize}
\item {\it \ttbar\ and Wt production:} \\
The dominant background contribution comes from $\ttbar$ 
production. 
The production cross-section is large and the leptonic 
decays (into $ e, \mu$ and $\tau $)
of the $W$'s, produced in the $t \rightarrow W b$ 
decays, lead to a signature similar to the signal. 
Due to the appearence of two b-jets that can mimic the forward jet tags, 
$\ttbar$ events contribute to the background already 
at leading order.

Another important background results from single-top production in 
association with a $W$ boson, where the leptons result from the $W$ 
and the top quark decay.

\item {\em QCD WW + jet production:} \\
The continuum production of $W$ pairs in association with two or more jets 
is another background.
The diagrams referred to as QCD-WW production involve
color exchange between the initial and final partons.

\item  {\em Electroweak WW + jet production:} \\
Pairs of $W$ bosons may also be produced in an electroweak (EW) process 
via vector boson exchange. Although this background has a much 
lower cross-section than the QCD-WW production, it shows similar 
characteristics to the signal process. The 
electroweak bosons exchanged in the $t$-channel are radiated off the initial 
quarks and no color flow is exchanged. 
Due to this similarity to the signal process, the rejection of this particular
background is expected to be much harder than for the QCD-type backgrounds. 

\item {\em QCD $\gamma^*/Z$ + jet production:}\\
The Drell-Yan $\gamma^*/Z$ + jet production has a large cross-section at the LHC. 
For electron and muon pairs in the final state, 
the main discriminating variable between signal and background is
the missing transverse momentum. In addition, the 
Z-resonance peak can be vetoed by applying a cut on the invariant mass of 
the same flavour di-lepton pair. 

Tau pairs in the final state from  $\gamma^*/Z \rightarrow \tau^+ \tau^-$
represent a potentially serious background for the Higgs search in the $\tau \tau$ 
decay mode. If both $\tau$-leptons decay 
leptonically, this process also contributes to the 
$\hwws \rightarrow \ell \nu \ell \nu$ search. 

In addition to Drell-Yan, processes involving gluon splitting in the 
initial state $qg \to qg \tau \tau, gg \to q \bar{q} \tau \tau$ or quark scattering,
dominated by $t$-channel gluon exchange with $Z$ or $\gamma$ bremsstrahlung, 
contribute to this background process.

\item {\em Electroweak $\tau \tau $ + jet production:}\\
Tau pairs can also be produced 
in an electroweak process, via a $t$-channel weak boson exchange. As in the 
case of the electroweak $WW$ production, a larger acceptance compensates the
smaller production cross-section.

\item {\em ZZ production:} \\
An additional background comes from ZZ production. There is 
a contribution where one Z decays into an $ee / \mu \mu$ pair and the other
hadronically, leading to jets. Another contribution results from $ZZ$ 
events where the second $Z$ decays into a pair of neutrinos and jets 
are produced from additional QCD radiation. 
\end{itemize}

An additional background that has not been considered in detail can arise from 
$\bbbar$ production, which has a huge cross-section at the LHC. Previous 
studies \cite{costanzo_1} have shown that requiring two high-$\PT$ 
isolated leptons in 
association with $\ptmiss$ and/or angular cuts on the leptons will   
suppress this background well below the level of the other backgrounds listed 
above. 
\begin{table}[h]
\footnotesize
\begin{center}
\begin{tabular}{l|l|r}
\hline
\hline
process & $P_T$-cutoff& cross-section  \\
\hline
\hline
$\gamma^*/Z  \ + jets $, $\gamma^*/Z \rightarrow \ell \ell $ & 
$ > 10$ GeV/$c$& 5227  pb \\
\ttbar   & & 55.0 pb \\
\qcdww   & & 16.7 pb \\
$q g \rightarrow Wt$ & & 4.8 pb \\
\hline
\ewtau   & & 170.8 fb \\
\ewww    & & 81.6 fb \\
\hline
\hline
\end{tabular}
\vspace{0.5cm}
\caption{\small \it cross-sections times 
leptonic branching ratios ($W \rightarrow \ell \nu$, 
$\ell = e, \mu $ and $\tau$) for the major background processes. 
}
\label{t:backgr}
\end{center}
\end{table}

The signal process and all background processes except the single top, the 
electroweak WW and $\tau \tau$ production have been generated 
using the PYTHIA 6.1 Monte Carlo 
event generator \cite{pythia}.  Initial- and 
final-state radiation (ISR and FSR) and fragmentation, as well as multiple 
interactions, have been switched on, thereby
allowing for a study of the 
jet activity in the central detector region due to radiation.
The CTEQ5L parametrization \cite{cteq-sf} of the parton distribution functions 
has been used in the generation of all signal and background processes.
To take the spin correlations properly into account, tau decays
have been modelled using the TAUOLA decay library \cite{tauola}.
The two electroweak processes, which are not included in PYTHIA, and the 
QCD $\gamma^*/Z+jet$ production have been 
generated by interfacing the parton-level generation of Ref.~\cite{zeppenfeld-ww}
to PYTHIA, which was then used to perform the parton showering, including 
initial- and final-state radiation~\cite{rachid}. 
The single-top background has been generated using the Monte Carlo programme 
of Ref.~\cite{onetop}.

Since the $\ttbar$ and the $\gamma^*/Z+jet$ with $Z \rightarrow \tau \tau$ 
backgrounds are the dominant backgrounds for the $WW^{(*)}$ and $\tau \tau$ 
decay modes respectively, 
these backgrounds have been generated also using explicit 
tree-level matrix element calculations for $\ttbar$ + 0, 1, and 2 
jets\footnote{The matrix element calculations were provided by the authors of 
Ref.~\cite{zeppenfeld-ww}} and 
$Z$ + 2 jets. The latter process has also been generated using the {\em COMPHEP}
Monte Carlo generator~\cite{comphep}. 

A summary of the major background processes and the relevant cross-sections 
multiplied by the branching ratios $BR(W \rightarrow \ell \nu)$ and 
$BR(Z \rightarrow \ell \ell )$, where $\ell=e,\mu$ and 
$\tau$ are listed in Table \ref{t:backgr}.  

ATLFAST \cite{atlfast}, the package for the fast simulation of the ATLAS 
detector, has been used to perform 
the detector simulation for all processes considered. This simulation provides
a parametrized response of the crucial detector
performance figures, based on detailed GEANT simulations \cite{atlas-tdr}.

\newpage  

\section{Experimental Issues}
In this section, several experimental issues that are common to all vector boson 
fusion analyses are discussed. Among them are the questions of the trigger, the 
forward jet tagging and the central jet veto. 

\subsection{Trigger aspects}
All channels considered have leptons ($e$ or $\mu$) in the final state and can be
triggered by either the single- or the di-lepton trigger. The hardest
trigger requirements are set by the $\tau \tau$ decay mode, where in
the di-lepton channel low
\PT-thresholds for the reconstructed leptons are needed to keep the
acceptance at a high level. Assuming the ATLAS trigger
thresholds \cite{atlas-trigger}, high trigger efficiency can be reached for a single
electron or muon for \PT\ values above 25~\Gc\ or 20~\Gc\ respectively. The 
ATLAS trigger acceptance covers the region $| \eta | <$ 2.5 for electrons and
$| \eta | <$ 2.2 for muons. The \PT-threshold values for the lepton pair triggers are 15~\Gc\ (for $ee$) and 
10~\Gc\ (for $\mu \mu$). For the mixed mode ($e , \mu$), it is assumed that each
lepton passes the respective threshold of 10~\Gc\ for muons and
15~\Gc\ for electrons or that at least one of them passes the higher 
thresholds of the single-lepton trigger. 
Lepton-hadron $\tau \tau$ final states can be triggered via the single-lepton 
trigger or via the hadronic tau + $\ptmiss$ trigger.

\subsection{Lepton Identification}
The reconstruction of electrons and muons has been studied in detailed detector 
simulations~\cite{atlas-tdr} including the effects of pileup. Based on these studies, 
the efficiency for the reconstruction of electrons or muons in the pseudorapidity range 
of the ATLAS Inner Detector, $|\eta | < 2.5$, is taken to be 90\%.

It is further assumed that hadronically decaying taus can be identified 
over the
same range of pseudorapidity. Important ingredients in the tau
identification~\cite{donatella} are the profile of the energy deposition 
in the calorimeter and the number of tracks with
a transverse momentum above 1~\Gc\ pointing to the calorimeter cluster. 
The tau reconstruction efficiency
is correlated with the rejection against jets~\cite{atlas-tdr,donatella}.
In the present study,
an efficiency of 50\% for hadronic tau decays is assumed. This leads
to a typical rejection~\cite{atlas-tdr} of about 100 for jets with 
a \PT\ of 40~\Gc.

\subsection{Jet Tagging}
From the production process it is expected that the two tag jets are 
reconstructed with a sizeable \PT\ in opposite hemispheres and have a 
large separation in pseudorapidity. Without further hard 
initial- or final-state radiation, the transverse momentum of the tagging jets 
should be balanced by the transverse momentum of the Higgs boson. 

\begin{figure}[hbtn]
\begin{center}
\begin{minipage}{7.7cm}
\mbox{\epsfig{file=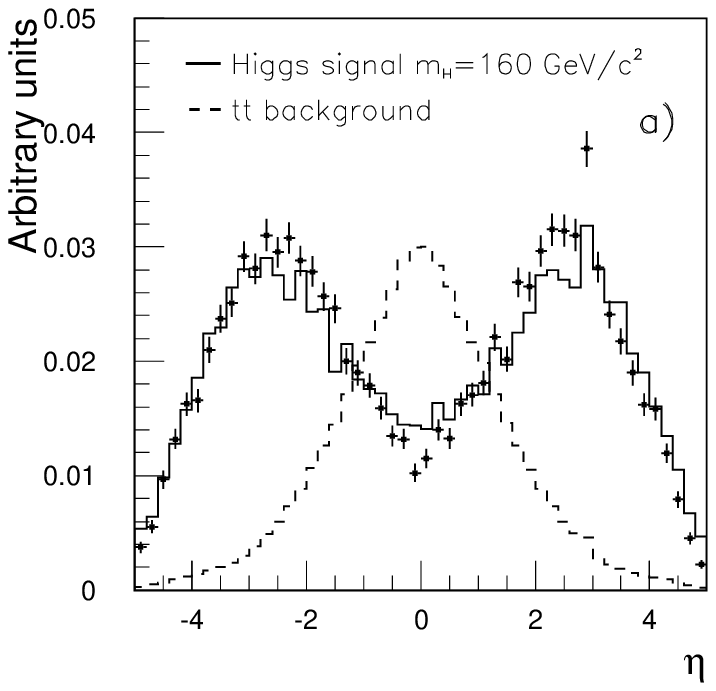,height=7.5cm}}
\end{minipage}
\begin{minipage}{7.7cm}
\mbox{\epsfig{file=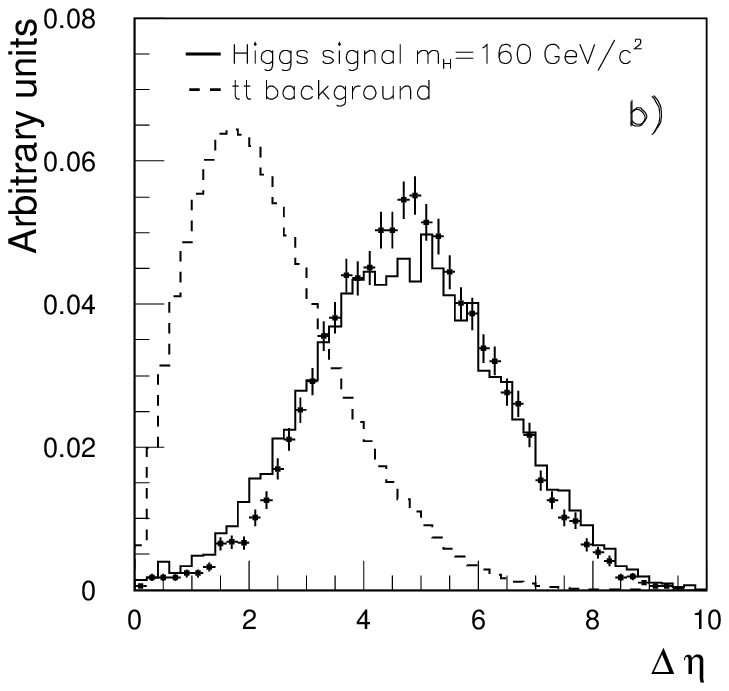,height=7.5cm}}
\end{minipage}
\caption{\small \it a) Pseudorapidity distribution of the tag jets in signal 
events with $m_H = 160$~\Gcs\ and for \ttbar\ background events. 
The full histograms show the distributions at parton level, the dots 
represent the reconstructed distributions, after the tagging algorithm 
has been applied. The corresponding distributions for jets identified as tag 
jets in \ttbar\ events are superimposed as dashed histograms. All distributions 
are normalized to unity. b) Separation $\Delta \eta$ between the tag jets
for the same types of events.}
\label{f:tagjets}
\end{center}
\end{figure}

In the present study, the two tag jets are searched for over the full 
calorimeter coverage of the ATLAS detector $(| \eta | < 4.9)$. For all
jets, a calibration has been applied which corrects the jet energy on average 
back to the original parton energy.  After the calibration, the jets with
the highest \PT\ in the positive and negative regions of pseudorapidity
are taken to be the tag jet candidates. 
Studies have shown \cite{jakobs01} that this choice of the 
tag jets has a high efficiency for correctly identifying the tag jets. 

The pseudorapidity distribution and the separation $\Delta \eta$ between 
the two tag jets, as found from the parton-level information, 
is shown in Fig.~\ref{f:tagjets}
for signal events with $m_H =$  160~\Gcs\ (full histogram). 
Using the tagging algorithm described above, good agreement is found 
between the parton level and the reconstructed information, which is 
represented by the dots in the figure. 
Due to the opposite hemisphere requirement, a small bias is 
found for jets with a small pseudorapidity separation; however, this does not
affect the final result. Also 
superimposed onto this figure are the corresponding distributions for 
tag jets as reconstructed in \ttbar\ background events. 
A comparison between the distribution for signal and background events 
clearly suggests that 
a large pseudorapidity separation should be used for the discrimination between 
signal and QCD-type backgrounds.

In order to address the question how well these jets can be identified 
at the LHC in the presence of pileup, a full GEANT 
simulation of the performance of the ATLAS detector has been 
performed~\cite{pisa-tagging}. This 
study has demonstrated that tag jets can be reliably reconstructed 
in the ATLAS detector. The efficiency for reconstructing a tag jet 
in signal events with 
\PT\ above 20 \Gc\ (originating from a parton with $\PT > 20$~\Gc) is shown 
in Fig.~\ref{f:jetveto}a as a function of pseudorapidity $\eta$.
The fast simulation package of the ATLAS 
detector provides a sufficiently good description of the tagging efficiency. 
Differences between the fast and full simulation have been found in the 
transition regions between different calorimeters and at very forward 
rapidities. These differences have
been parametrized as a function of \PT\ and $\eta$ and have been used
to correct the fast simulation results accordingly~\cite{pisa-tagging}. 

\subsection{Jet-Veto Efficiencies}
\begin{figure}[hbtn]
\begin{center}
\begin{minipage}{7.5cm}
\mbox{\epsfig{file=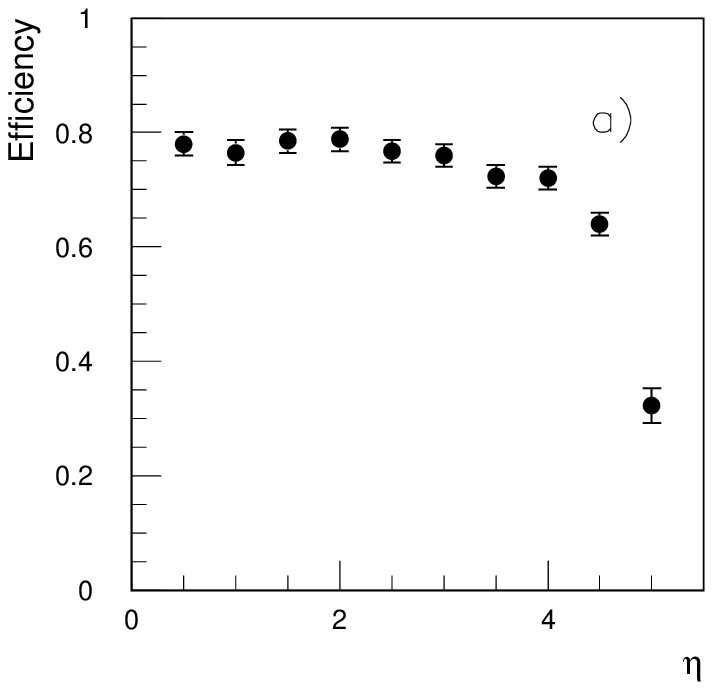,height=7.5cm}}
\end{minipage}
\begin{minipage}{7.5cm}
\mbox{\epsfig{file=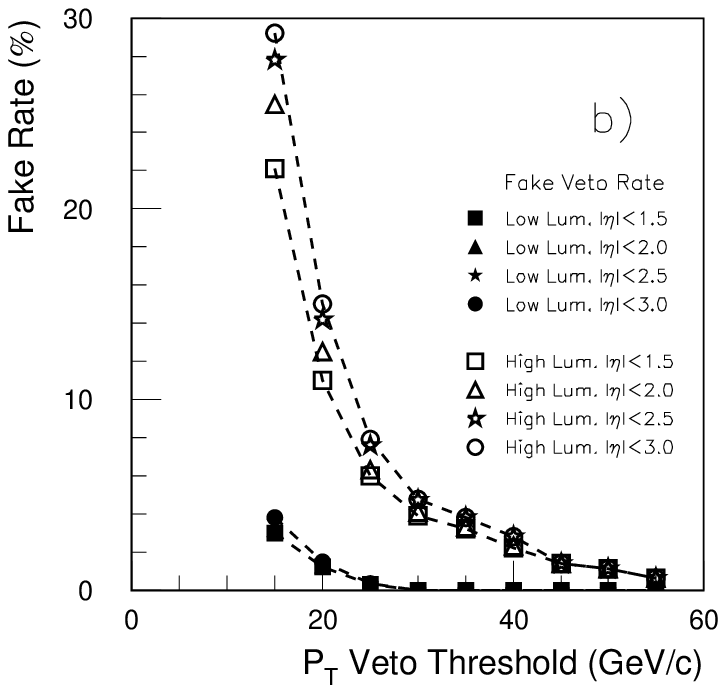,height=7.5cm}}
\end{minipage}

\caption{\small \it 
a) Efficiency for reconstructing a tag jet with $\PT > 20$~\Gc\ which originates
from a parton with $\PT > 20$~\Gc\ as a function of pseudorapidity $\eta$
of the parton.
b) Probability for finding at least one jet from pileup events in 
central rapidity intervals
in the ATLAS detector as a function of the \PT\ value used in the jet definition. 
The dashed curves connect the points for pseudorapidity itervals 
$| \eta |<$1.5 and $| \eta |<$3.0 for low and high LHC luminosities.}
\label{f:jetveto}
\end{center}
\end{figure}

As pointed out above, a veto against jets in the central region
will be an important tool to suppress QCD backgrounds. At the LHC, jets in the 
central region can be produced also by pileup events. In the full simulation 
study~\cite{pisa-tagging}, it has been found that after applying a
threshold cut on the calorimeter
cell energies of 0.2~GeV at low and 1.0~GeV at high luminosity, the fake 
jets from pileup events can be kept at a low level, provided that \PT\
thresholds of 20~\Gc\ at low and 30~\Gc\ at high luminosity are used for the 
jet definition. The results of this study are presented in 
Fig.~\ref{f:jetveto}b,
where the efficiency to find a jet from pileup events in different 
intervals of central rapidity is 
shown as a function of the jet \PT-threshold for low and high luminosity. This 
behaviour demonstrates that at high luminosity, a jet \PT-threshold
of 30~\Gc\ or higher should be used.

\section{Signal Selection Criteria}

Similar signal characteristics and reconstruction methods are
exploited in the various channels considered. Common features are
discussed in the following: 

\begin{itemize}
\item {\em Lepton Cuts:} \\
A characteristic feature of the $WW^{(*)}$ decay channel is the 
anti-correlation
of the W spins from the decay of the scalar Higgs boson \cite{97ditt}.
Since the $W^+$ and $W^-$ have opposite spins, the lepton and anti-lepton
tend to be emitted in the same direction.
A series of cuts is performed on the angular separation of the charged
leptons, namely the azimuthal angle $\Delta \phi_{\ell \ell}$ between the lepton 
directions,
the cosine of the polar opening angle $\cos \theta_{\ell \ell}$ and the separation
in $\eta - \phi$ space $\Delta R_{\ell \ell}$. 
Since in the rest frame of the Higgs boson, the di-lepton system and the 
neutrino system are emitted back-to-back with equal energy, the invariant
mass of the visible leptons, $M_{\ell \ell}$, is limited to $\sim m_H / 2$. 
A cut on $M_{\ell \ell}$ can therefore be applied.
Finally, cuts on the maximum lepton transverse momentum can be applied to 
reject background events without a significant loss of signal efficiency.
Distributions for the variables $\Delta \phi_{\ell \ell}$ and $M_{\ell \ell}$
are shown in Fig.~\ref{f:ww-lep} for signal events with $m_H = 160$~\Gcs\
and for events from several background sources.

\begin{figure}[hbtn]
\begin{center}
\begin{minipage}{7.7cm}
\mbox{\epsfig{file=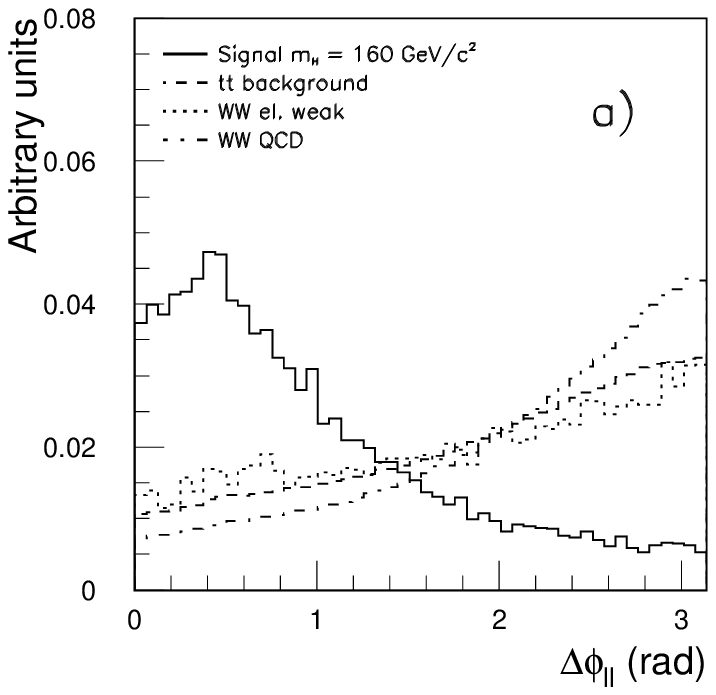,height=7.5cm}}
\end{minipage}
\begin{minipage}{7.7cm}
\mbox{\epsfig{file=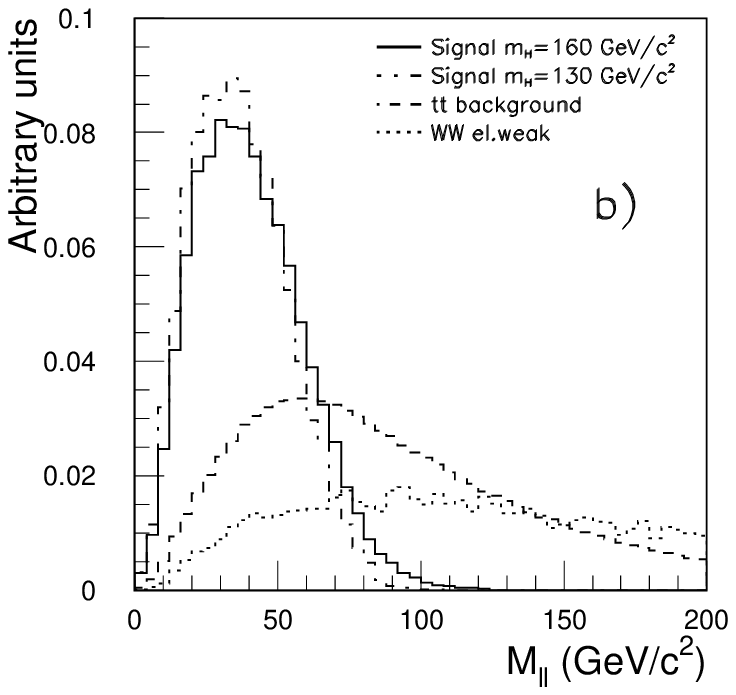,height=7.5cm}}
\end{minipage}
\caption{\small \it 
a) Azimuthal angle separation $\Delta \phi_{ll}$ between the leptons from signal 
events with $m_H = 160$~\Gcs\ and different backgrounds.
b) Di-lepton invariant mass distribution
for signal events with $m_H = 160$~\Gcs\ and $m_H = 130$~\Gcs, 
for \ttbar\ and for electroweak WW background. 
All distributions are normalized to unity.}
\label{f:ww-lep}
\end{center}
\end{figure}

\item{\em Tag jets:} \\
The two tag jets are required to pass  \PT-thresholds and to have a
minimum separation in pseudorapidity
$\Delta\eta_{tags} = \eta_{tag}^{max} - \eta_{tag}^{min}$.  
In addition, it is required that the identified leptons be reconstructed in 
the central region of the detector and lie in the pseudorapidity gap spanned
by the two tag jets. Tag jets identified in
QCD processes typically have smaller invariant jet-jet masses than those 
identified in electroweak processes. Therefore, the invariant mass 
$m_{jj}$ of the tagged jets provides a selection criterion for the 
signal process.

If no hard initial- or final-state gluons are radiated, it is also 
expected that the transverse momentum of the Higgs boson will be balanced 
by the transverse momentum of the two tag jets. This behaviour is illustrated
in Fig.~\ref{f:pt_bal}, where the modulus of the vector  

\[
\vec{P}_T^{tot} = \vec{P}_T^{\ell,1} + \vec{P}_T^{\ell,2} + \vec{P}_T^{miss} 
                + \vec{P}_T^{j,1} + \vec{P}_T^{j,2} 
\]

is shown for \hwwll\ signal events and for \ttbar\ and QCD-$WW$ background 
events. A cut on the variable $|\vec{P}_T^{tot} | $ is
correlated with a jet-veto cut, if the jet which triggers the 
veto originates from the hard-scattering process in which the Higgs 
boson is produced. However, this variable is largely insensitive to 
jets which result from pileup effects in the detector. Therefore, a 
useful rejection can be obtained, even at high LHC luminosity. The 
only degradation effect expected at high luminosity is a broadening
of the  $|\vec{P}_T^{tot} |$ spectrum due to the degraded \ptmiss\ resolution. 

\begin{figure}[hbtn]
\begin{center}
\mbox{\epsfig{file=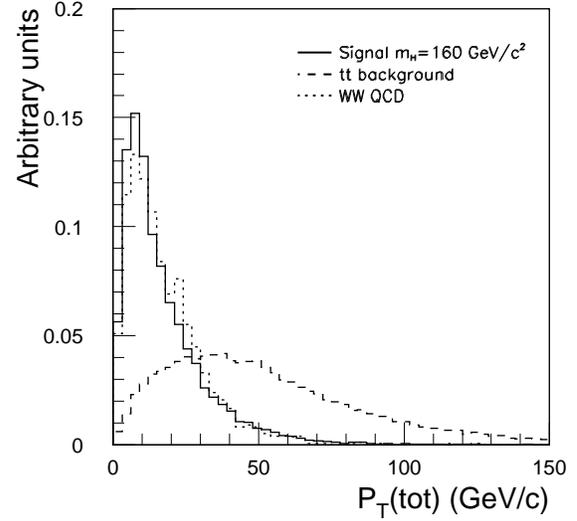,height=7.5cm}}
\caption{\small \it 
Distribution of the momentum balance $|\vec{P}_T^{tot} |$ 
between the reconstructed leptons, \ptmiss\ and the tag jets.
All distributions are normalized to unity. }
\label{f:pt_bal}
\end{center}
\end{figure}

\begin{figure}[hbtn]
\begin{center}
\mbox{\epsfig{file=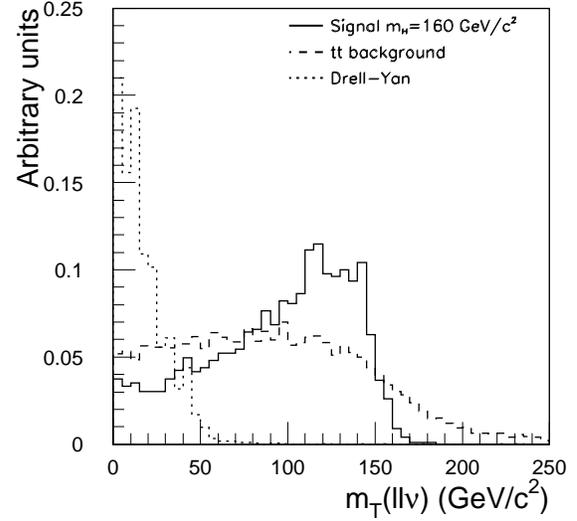,height=7.5cm}}
\caption{\small \it 
Distribution of the transverse mass $m_T(ll\nu)$ of the di-lepton and neutrino 
system for signal events with $m_H = $ 160~\Gcs\ and for 
different backgrounds. The distributions are normalized to unity.}
\label{f:mt_llv}
\end{center}
\end{figure}

\item{\em Central Jet veto:} \\
Events are rejected if at least one jet with a transverse momentum 
above 20~\Gc\ is found in the pseudorapidity range spanned
by the two tag jets. For those channels where \ttbar\ is a significant 
background, the jet-veto region has been enlarged to cover the pseudorapidity
range $-3.2 < \eta < 3.2$. Using this rapidity window, which is usually larger 
than the gap between the tag jets, events
with high jet multiplicities (like \ttbar\ events due to the additional b-jets)
are better rejected. 

\item {\em Drell-Yan rejection:} \\
For those processes where same flavour leptons are considered in the 
final state the Drell-Yan production represents a serious background. This 
background also exists, although with a lower rate,  for $e$-$\mu$ 
final states via the Drell-Yan production of tau pairs with subsequent
double leptonic decays. The $Z\rightarrow ee / \mu \mu$ components can be 
largely rejected by vetoing those events 
which have an invariant di-lepton mass compatible with the Z-mass.
An efficient rejection of background contributions at low mass can 
be achieved via cuts on the missing transverse momentum or on 
the reconstructed 
transverse mass $m_T (\ell \ell \nu) $ of the di-lepton and neutrino
system, which is defined as 
$m_T (\ell \ell \nu) = \sqrt{2 P_T(\ell \ell) \ptmiss \cdot ( 1 - cos \Delta \phi )}$
where $\Delta \phi$ is the angle between the di-lepton vector and the \ptmiss\ 
vector in the transverse plane. The distributions of this variable for
signal,  \ttbar\ and  Drell-Yan background ($\gamma^* / Z + jets,
Z \rightarrow \tau \tau$) are shown in Fig.~\ref{f:mt_llv}.

\item{\em Tau reconstruction:} \\
For the $\htautau$ channel, the reconstruction of 
the $\tau$ momenta and thereby the Higgs boson mass is 
important. In the case of the \hwws\ decay mode, the tau reconstruction is
important for the rejection of the $\gamma^{*}/Z$ +jet 
(with $Z \rightarrow \tau \tau$) background.
 
In signal and background events, the $H$ or $Z$ bosons are emitted with 
quite high \PT\,
which contributes to large tau boosts and causes the tau decay products 
to be nearly collinear in the laboratory frame. 
Within the collinear approximation, {\em i.e.,} assuming that the tau directions 
are given by the directions of the visible tau decay products
(leptons or hadronic tau respectively), the tau momenta can be 
reconstructed. Labeling by $x_{\tau_1}$
and $x_{\tau_2}$ the fraction of the tau energy carried by each lepton
or hadronic tau system, the 
missing transverse momentum vector can be used to solve the two equations for 
the two unknowns  $x_{\tau_1}$ and $x_{\tau_2}$. For $\tau$ decays, they
should lie in the interval $ 0 < x_{\tau_{1,2}} < 1$. 
Distributions of the reconstructed variables $x_{\tau_1}$ versus $x_{\tau_2}$
are shown in Fig.~\ref{f:xtau} for \htautau\ signal events and 
background events from electroweak WW production. 

\begin{figure}[hbtn]
\begin{center}
\begin{minipage}{7.7cm}
\epsfig{file=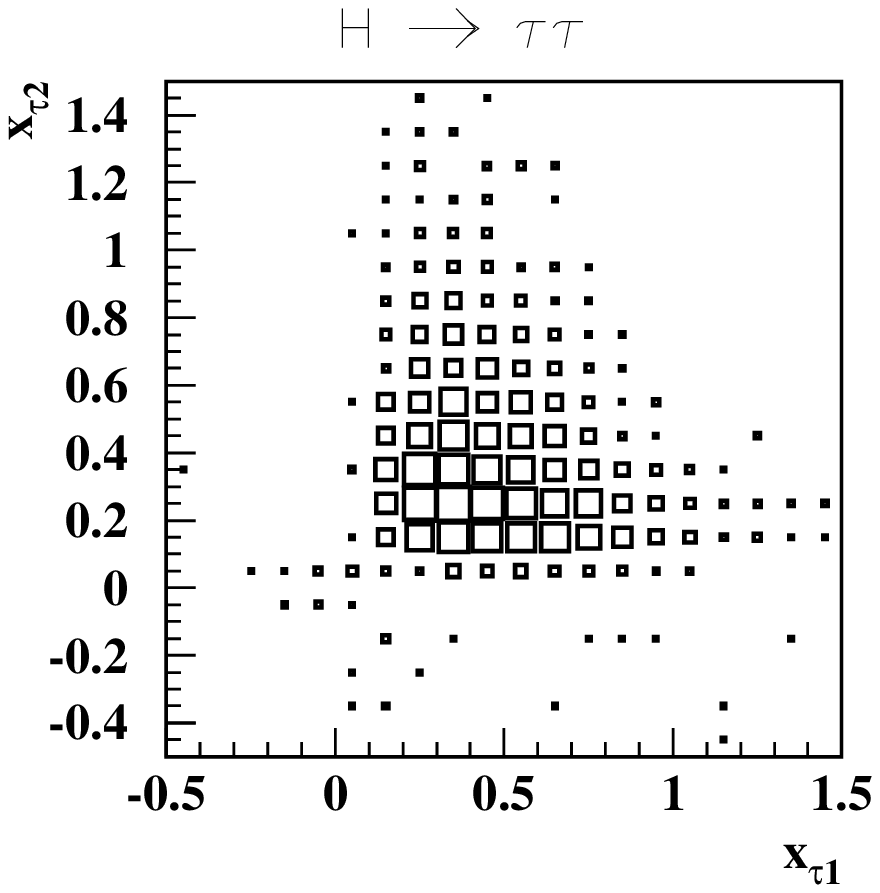,height=5.5cm}
\end{minipage}
\begin{minipage}{7.7cm}
\mbox{\epsfig{file=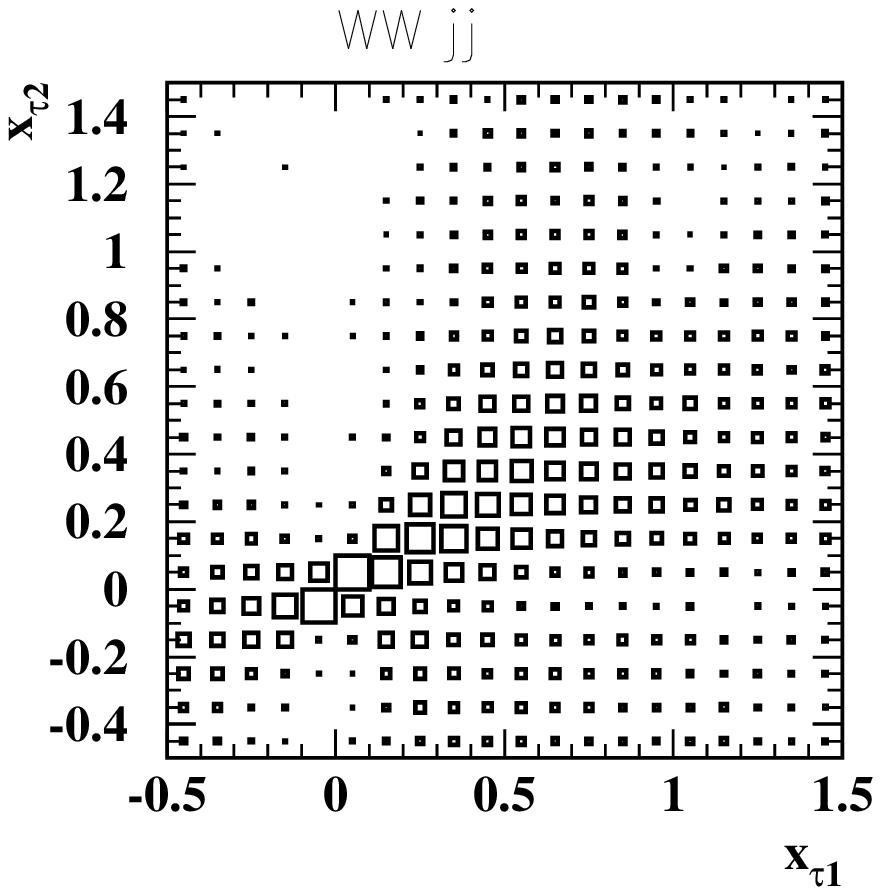,height=5.5cm}}
\end{minipage}
\caption{\small \it 
Distribution of $x_{\tau_1}$ versus $x_{\tau_2}$ for  \htautau\ signal 
events
with $m_H = 120$~\Gcs\ (top) and for events from electroweak WW
production (bottom).}
\label{f:xtau}
\end{center}
\end{figure}

\item{\em Mass reconstruction:} \\
In the case of the $\tau \tau$ decay modes, the Higgs boson mass 
is given by 

\[ 
m_{\tau \tau} = m_{\ell \ell } / \sqrt{x_{\tau_1}x_{\tau_2}}. 
\]

In the \hwws\ decay mode, a direct reconstruction of the Higgs boson mass 
is not possible, since the longitudinal component of the momentum of the
neutrino system can not be reconstructed. 
For masses below 2$m_W$, the (virtual) $W$
bosons are at mostly at rest in the Higgs boson center-of-mass system, 
resulting in 
$m_{\ell \ell} \approx m_{\nu \nu}$, so that for both the di-lepton and for the 
neutrino system the transverse energy can be calculated as \cite{zeppenfeld-ww}

\[ 
E_T^{\ell \ell} = \sqrt{(P_T^{\ell \ell})^2 + m^2_{\ell \ell}}, 
\]
\[
E_T^{\nu \nu} = \sqrt{(\ptmiss)^2  + m^2_{\ell \ell}} . 
\]
Using these transverse energies the transverse mass of the di-lepton-\ptmiss\
system can be calculated as

\[ 
M_T = \sqrt{(E_T^{\ell \ell} + E_T^{\nu \nu})^2 - ( \vec{p}_T^{\ \ell \ell} + 
\vec{p}_T^{\ miss} )^2 }.
\]

For Higgs boson masses up to $\sim$ 160~\Gcs, this is a good approximation
for the Higgs boson transverse mass. Also,
above the threshold for the decay into two real $W$'s, the reconstructed
mass still contains information about the Higgs boson mass. Again it is
found that this definition leads to a sharper signal peak than the
previously defined $m_T (\ell \ell \nu )$, where the di-lepton mass has
been neglected. 
\end{itemize}

\section{The \hwws\ decay mode}
In this Section the detailed analysis is described for the \hwws\ 
channels. An optimisation of the cuts has been performed in the mass
region $160$~\Gcs $< m_H <$ 180~\Gcs and for low masses around 
120~\Gcs. Details can be found in Refs.~\cite{jakobs01} and \cite{mellado} 
respectively.

\subsection{Di-lepton final states: $\hwwll$ }

\subsubsection{Event selection} 
In the event selection, the cuts listed in Table~\ref{t:wwcuts} have been 
applied in the two different mass regions.  
\begin{table*}
\begin{center}
\begin{minipage}{0.75\linewidth}
\footnotesize
\begin{center}
\begin{tabular}{ |l|c|c| }
\hline
\hline
cut        & high mass & low mass \\
           & 135~\Gcs\ $ < m_H <$ 190~\Gcs\  &  110~\Gcs\ $ < m_H <$ 135~\Gcs\ \\ 
\hline
\hline
Two leptons with       
& $ P_{T}^1 > 20$~\Gc &  $ P_{T}^1(e) > 15$ \Gc  \\
& $ P_{T}^2 > 15$~\Gc &  $ P_{T}^1(\mu) > 10$ \Gc \\
&                     &  $ P_{T}^2(e)  > 15$ \Gc  \\
&                     &  $ P_{T}^2(\mu) > 10$ \Gc\ \\
& $| \eta| < 2.5$     &  $| \eta| < 2.5 $    \\
\hline
Tag jets 
&  \multicolumn{2}{c|}{$P_T^1 > 40 $ \Gc, $P_T^2 > 20 $\Gc} \\
&  \multicolumn{2}{c|}{$ \Delta \eta_{tags} = |\eta_{tag}^{1} - \eta_{tag}^{2} | > 3.8 $} \\
Leptons between tag jets  
& \multicolumn{2}{c|}
    {$\eta_{tag}^{min} <  \eta_{l_{1,2}}  < \eta_{tag}^{max}$}\\
\hline
Lepton Cuts & $\Delta \phi_{\ell \ell} \le 1.05$ 
            & $\Delta \phi_{\ell \ell} \le 1.5$ \\
            & $\Delta R_{\ell \ell} \le 1.8$ 
            & $\Delta R_{\ell \ell} \le 1.6$ \\
            & $\cos \theta_{\ell \ell} \ge 0.2$  & \\
            & $M_{\ell \ell} <$ 85~\Gcs\      
            & $M_{\ell \ell} <$ 65~\Gcs\  \\               
            & $P_{T} (\ell_{1,2}) <$ 120~\Gc    & \\
\hline
Tau veto    & \multicolumn{2}{c|}{reject events if 
$ x_{\tau_{1}}, x_{\tau_{2}} > 0$} \\
            & \multicolumn{2}{c|}
{$ | M_{\tau \tau} -  M_{Z} |  <  25$~\Gcs} \\
\hline
Invariant mass of the & $M_{jj} >$ 550~\Gcs\ 
                                   & 600~\Gcs\ \ $< M_{jj} \ <$ 2500~\Gcs \\
two tag jets & & \\
\hline
Transverse momentum & \multicolumn{2}{c|}
{$ | \vec{P}_T^{tot} | < 30~\Gc$}  \\ 
balance & \multicolumn{2}{c|}{  \ } \\
\hline
Jet veto   &\multicolumn{2}{c|}{no jets with $\PT > 20$~\Gc\ in  $| \eta | < 3.2$}\\
\hline
$\gamma^{*} /Z, Z \rightarrow \tau \tau$ rejection: 
& $m_T (\ell \ell \nu) > 30$ \Gcs\  
& $m_T (\ell \ell \nu) > 20$ \Gcs\    \\
\hline
\hline
\end{tabular}
\vspace{0.5cm}
\caption{\small  \it Cuts applied in the \hwws\ analyses at high and
low masses.}
\label{t:wwcuts}
\end{center}
\end{minipage}
\end{center}
\end{table*}
Since it is not expected in the signal that tag jets are b-jets, the selection
of tag jet candidates can be enhanced with b-tagging. For instance, in the present 
analysis \ttbar\ production, which constitutes the largest background, typically has 
one b-jet identified as a tagging jet. In order to suppress this background it is 
required that a tag jet candidate is not tagged as b-jet. This requirement can, 
however, only be applied for the tag jet candidates that fall in the acceptance 
region of the ATLAS Inner Detector, {\em i.e.}, $ | \eta | < 2.5$. 
The b-jet efficiency versus rejection against
non b-jets has been optimized. A maximum in signal significance
has been found for a b-jet efficiency of 0.70 and a corresponding 
mistag probability of 0.25 for c-quark jets and 0.04 for light quark 
and gluon jets.

The additional background contributions 
for the signal from same-flavour leptons, of which 
the $ee$ and $\mu \mu$ Drell-Yan backgrounds are the dominant ones, 
can be efficiently rejected by tightening the di-lepton mass cut and by 
introducing 
a \ptmiss\ cut:

\begin{itemize}
\item $ M_{\ell \ell} < 75$~\Gcs\   and 
\item $\ptmiss > 30$~\Gc\ 
\end{itemize}

The acceptance for a Higgs boson with a mass of 160 \Gcs\ and for the 
backgrounds after the application of successive cuts 
is summarized for the $e \mu$ final state in detail in Table~\ref{t:ww-acc}. 
Also the contributions to the signal from
other Higgs boson production processes have been considered.
A significant contribution has been found
to arise from the dominant gluon-gluon fusion process $gg \rightarrow 
\hwws $ where the two tag jets are produced from initial- and 
final-state radiation. Contributions from the associated production processes
$W H, ZH $ or $\ttbar H$, where leptons come either from the Higgs 
boson decays via $WW^{(*)}$ or  from the decays of the accompanying vector
boson or top-quarks, have been found to be small, with a final accepted 
cross-section below 0.05 fb and have been neglected.  

\renewcommand{\baselinestretch}{1.1}\selectfont
\begin{table*}
\begin{center}
\begin{minipage}{0.75\linewidth}
\footnotesize 
\begin{center}
\begin{tabular}{l||r r||c|r|r|r|r|r}
\hline
\hline
 & \multicolumn{2}{c||}{signal (fb)} & \multicolumn{6}{c}{background (fb)} \\
 &VV & gg & $\ttbar \ + \ Wt$ & \multicolumn{2}{c|}{\wwjets} & \multicolumn{2}{c|}{
$\gamma^*/Z + jets$} & total \\
 & & & & EW & QCD & EW & QCD & \\
\hline
Lepton acceptance       & 29.6 & 121.9 & 6073 & 14.2 &590.5 & 5.96 &25222& 31906 \\
+ Forward Tagging       & 11.4 &  2.24 &127.5 & 8.01 & 1.41 & 1.55 &208.3& 346.8 \\
+ Lepton cuts           & 6.95 &  1.36 & 17.0 & 0.54 & 0.17 & 0.50 & 30.0&  48.2 \\
+ $\tau$ rejection      & 6.64 &  1.34 & 16.3 & 0.50 & 0.17 & 0.09 & 5.93&  23.0 \\
+ Jet mass              & 5.30 &  0.76 & 10.0 & 0.50 & 0.06 & 0.09 & 4.01&  14.6 \\
+ $P_T^{tot}$           & 4.52 &  0.50 & 2.34 & 0.38 & 0.04 & 0.07 & 2.70&  5.53 \\
+ Jet veto              & 3.87 &  0.34 & 0.72 & 0.34 & 0.03 & 0.07 & 1.70&  2.86 \\
+ $m_T(\ell \ell \nu)$ cut & 3.76 &  0.31 & 0.66 & 0.32 & 0.02 & 0.01 & 0.03&  1.04 \\
\hline 
\hline
$\hwws \rightarrow e \mu + X $&    &       &      &      &      &      &     &      \\
incl. $\tau \rightarrow e, \mu$ contr. 
                        & 4.32 & 0.33  & 0.75 & 0.35 & 0.03 & 0.01 & 0.03& 1.17 \\
\hline
$\hwws \rightarrow ee / \mu \mu + X$  &    &       &      &      &      &      &    &\\
                              
incl. $\tau \rightarrow e, \mu$ contr.
                        & 3.92 & 0.30 & 0.71 & 0.36 & 0.04 & 0.04 & 0.12& 1.27 \\
\hline
\hline
\end{tabular}
\vspace{0.5cm}
\caption{\small \it 
Accepted signal (for $m_H =$ 160~\Gcs) and background cross-sections 
 in fb for the $H \rightarrow WW \rightarrow e \mu \nu \nu$ 
channel after the application of 
successive cuts. For the signal, the contributions via the 
vector boson fusion (VV) and the gluon fusion channel  (gg) 
are given separately. The last two lines give the
 final numbers if the contributions from $W\rightarrow \tau \nu
 \rightarrow l \nu \nu \ \nu$ are added for both the $e \mu$ and the
 $ee / \mu\mu$ final states.}
\label{t:ww-acc}
\end{center}
\end{minipage}
\end{center}
\end{table*}

At the level of the basic acceptance cuts, {\em i.e.} requiring two leptons
and two tag jets with a pseudorapidity separation of
$\Delta \eta = 3.8$, the \ttbar\ and the $\gamma^*/Z + jet$ background 
are each about one order of magnitude larger than the signal. 
Due to the requirement of forward tag jets, the signal is already at that 
level dominated by contributions from the vector boson fusion subprocess. 
The lepton cuts suppress all backgrounds by nearly an order of
magnitude, whereas the signal is kept with an efficiency of about
65\%. Due to the tau reconstruction, a significant fraction of the 
electroweak and the $\gamma^*/Z + jet$ background can be rejected, 
whereas the 
acceptance for all other processes is high.
The cuts on the invariant mass of the tag jets and on the 
momentum balance significantly suppress all QCD
processes, in particular the dominant \ttbar\ background. 
The signal-to-background ratio can be further improved by applying the
central jet veto (mainly \ttbar\ rejection) and the 
cut on the transverse mass $m_T (ll \nu)$ (rejects mainly the $\gamma^*/Z$
background). 
After all cuts, the dominant backgrounds are \ttbar\ and $Wt$, due to their 
large 
production cross-sections, and electroweak $WW$ production, due to its 
similarity to the signal. For the top backgrounds, the \ttbar\ contribution is  
larger than the $Wt$ contribution. Out of the 0.66 fb quoted for the sum of the 
top backgrounds 
in the upper part of the Table, 0.51 fb are due to \ttbar\ production and 0.15 fb
result from single-top production. 

All numbers given in the upper part of Table~\ref{t:ww-acc}
come from direct decays with an  
electron and a muon in the final state. Di-leptons can, however, also be produced 
via cascade decays of tau leptons, for example, 
$ W \rightarrow \tau \nu \rightarrow \ \ell \nu \bar{\nu} \ \nu$.  
These contributions have also been 
calculated and have been added to the accepted signal and background 
cross-sections. An increase of about 15\% for the accepted cross-sections 
has been found.
Due to the softer \PT\ spectra of leptons
from tau decays this contribution is smaller than expected 
from a scaling of branching ratios. 

The final acceptance including the contributions from $\tau$ cascade
decays, is also given for the sum of 
the $ee$ and $\mu \mu$ final states. 
Due to the additional cuts the signal acceptance is  
lower than in the $e \mu$ case. However, the \ptmiss\ and the $M_{\ell \ell}$ 
cuts are very 
efficient in rejecting the additional $\gamma^* / Z + jet$ background, such that 
also in this channel, a convincing Higgs boson signal for Higgs boson masses 
around 160~\Gcs\ can be expected. 

After all cuts, a signal in the $e \mu$ channel
for a Higgs boson with a mass of 160~\Gcs\ of the order of 4.6 fb is 
found above a total background of 1.2 fb. 
Due to this large signal-to-background ratio, this channel alone has 
a good discovery potential for a Higgs boson with a mass  
around 160~\Gcs\ and is not very
sensitive to systematic uncertainties on the level of the background. 
This compares favourably with the  $gg \rightarrow WW^{(*)}$ channel previously
studied \cite{atlas-tdr,jakobs-ww}.
For an integrated luminosity of 30~\fbs\ and for a Higgs boson mass of 160~\Gcs, 
the background for the inclusive channel was found to be at a level of 650 events 
with a signal-to-background ratio of 0.60.

The corresponding numbers for a Higgs boson with a mass of 120~\Gcs\ are given 
in Table~\ref{t:ww-acc120}. Due to the suppressed branching 
ratio, the signal significance is reduced for lower Higgs boson masses. 
The contributions from cascade decays via tau leptons are found to be larger 
than in the high mass case, due to the lower \PT-thresholds used for the 
leptons. 
These contributions are included in the numbers given in Table~\ref{t:ww-acc120}.

\renewcommand{\baselinestretch}{1.1}\selectfont
\begin{table*}
\begin{center}
\begin{minipage}{0.75\linewidth}
\footnotesize 
\begin{center}
\begin{tabular}{l||r r||r|r|r|r|r|r}
\hline
\hline
 & \multicolumn{2}{c||}{signal (fb)} & \multicolumn{6}{c}{background (fb)} \\
 &VV & gg & $\ttbar \ + \ Wt$ & \multicolumn{2}{c|}{\wwjets} & \multicolumn{2}{c|}{
$\gamma^*/Z + jets$} & total \\
 & & & & EW & QCD & EW & QCD & \\
\hline
Lepton acceptance       & 5.20 & 17.30 & 8456 & 17.1 &617.2 & 7.09 & 4980& 14077 \\
+ Forward Tagging       & 1.85 &  0.27 & 82.6 & 10.7 & 1.83 & 2.10 & 45.2& 142.4 \\
+ Lepton angular cuts   & 1.36 &  0.18 & 13.5 & 0.89 & 0.27 & 0.81 & 7.47&  22.9 \\
+ $\tau$ rejection      & 1.27 &  0.18 & 12.9 & 0.83 & 0.27 & 0.15 & 1.64&  15.8 \\
+ Jet mass              & 0.88 &  0.08 & 6.39 & 0.43 & 0.08 & 0.11 & 0.83&  7.84 \\
+ $P_T^{tot}$           & 0.68 &  0.05 & 1.40 & 0.32 & 0.04 & 0.10 & 0.46&  2.32 \\
+ Jet veto              & 0.59 &  0.05 & 0.61 & 0.28 & 0.04 & 0.10 & 0.32&  1.35 \\
+ $m_T(\ell \ell \nu)$-cut & 0.52 &  0.05 & 0.58 & 0.27 & 0.03 & 0.02 & 0.05&  0.95 \\
\hline
\hline
$\hwws \rightarrow e \mu + X $
                        & 0.52 & 0.05  & 0.58 & 0.27 & 0.03 & 0.02 & 0.05& 0.95 \\
\hline
$\hwws \rightarrow ee / \mu \mu + X$ 
                        & 0.50 & 0.04 & 0.58 & 0.30 & 0.03 & 0.03 & 0.39 & 1.33 \\
\hline
\hline
\end{tabular}
\vspace{0.5cm}
\caption{\small \it 
Accepted signal (for $m_H =$ 120~\Gcs) and background cross-sections 
in fb for the $H \rightarrow WW \rightarrow e\mu \nu \nu $ channel after the application of 
successive cuts (upper part). The final numbers for both $e \mu$ and  
$ee / \mu \mu$ final states are given in the last two lines. 
The contributions from 
$W\rightarrow \tau \nu \rightarrow \ell \nu \nu \ \nu$ are included in all numbers
quoted. 
}
\label{t:ww-acc120}
\end{center}
\end{minipage}
\end{center}
\end{table*}

In the present study, numbers for signal and background have
been found which are somewhat different from the numbers quoted in the original 
parton-level study of Ref.~\cite{zeppenfeld-ww}. 
A detailed comparison between both simulations has been performed. 
If identical cuts as in Ref.~\cite{zeppenfeld-ww} are applied, the lepton acceptance 
is found to be about 20\% lower. In addition, 
the efficiency for reconstructing the tag jets has been found to be lower by 15\%. 
It has been traced back that both losses are related to initial- and final-state 
gluon radiation. They lead to 
a degraded lepton isolation as well as to non-Gaussian tails in the jet response
which can not be fully corrected in jet calibration procedures. However, the 
main conclusion of Ref.~\cite{zeppenfeld-ww}, 
that the search for vector boson fusion 
in the intermediate mass range at the LHC has a large discovery potential for 
a Standard Model Higgs boson in the \hwws\ decay channel, is confirmed by 
the present studies based on a more realistic detector simulation.

\subsubsection{Uncertainties on the background}

All QCD-type backgrounds 
quoted in the previous section have been calculated using
the PYTHIA parton-shower Monte Carlo. Due to the requirement of two hard 
tag jets in the final state, the final number of predicted background events is 
sensitive to the hard tail of the jet distribution in $\ttbar +  jet$ 
or $WW+jet$ events. 

In order to get an estimate on the systematic uncertainties on the background
predictions, the dominant $\ttbar + jet$ background has also been evaluated  
using explicit matrix element calculations for $\ttbar + 0 \ jet$,  
$\ttbar + 1 \ jet$ and 
$\ttbar + 2 \ jet$ final states. These matrix element
calculations have been provided by the authors of
Ref.~\cite{zeppenfeld-ww}
and have been interfaced to the PYTHIA generator. 
In order to avoid double 
counting when adding the three contributions, the procedure proposed in 
Ref.~\cite{zeppenfeld-ww}, to define three distinct final-state jet topologies, 
has been adopted. For $\ttbar \ + 0 \ jets$ only the two b-jets are considered
as tag jet candidates. Initial- and final-state radiation in these events may 
lead to a rejection of the event due to the jet veto. A distinctively different
class is defined by those $\ttbar \ + 1 \ jet$ events where the final-state 
light quark or 
gluon gives rise to one tag jet and one of the two b-jets is identified as the
other tag jet. Finally, a third class is defined where in $\ttbar \ + 2 \ jets$ 
the final-state light quarks or gluons are identified as tag jets. 

\begin{figure}[hbtn]
\begin{center}
\begin{minipage}{7.7cm}
\mbox{\epsfig{file=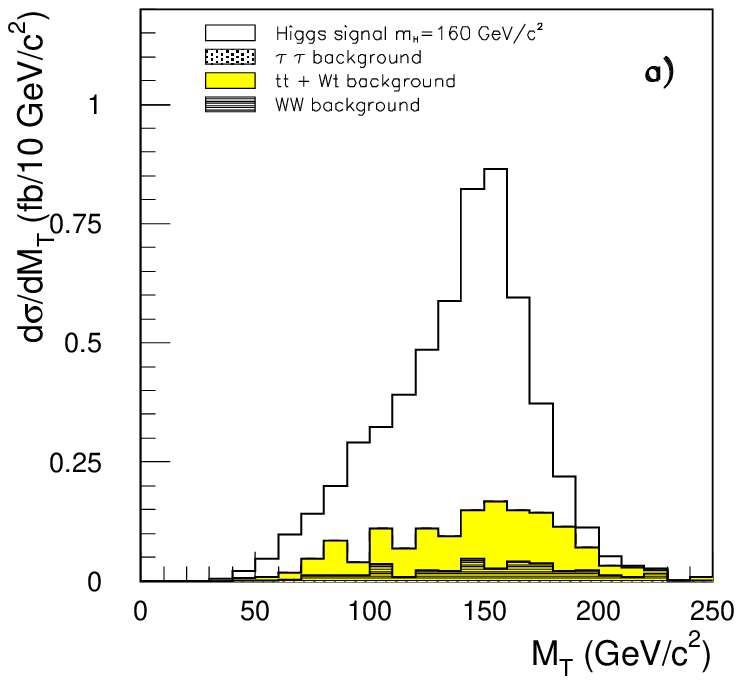,height=7.5cm}}
\end{minipage}
\begin{minipage}{7.7cm}
\mbox{\epsfig{file=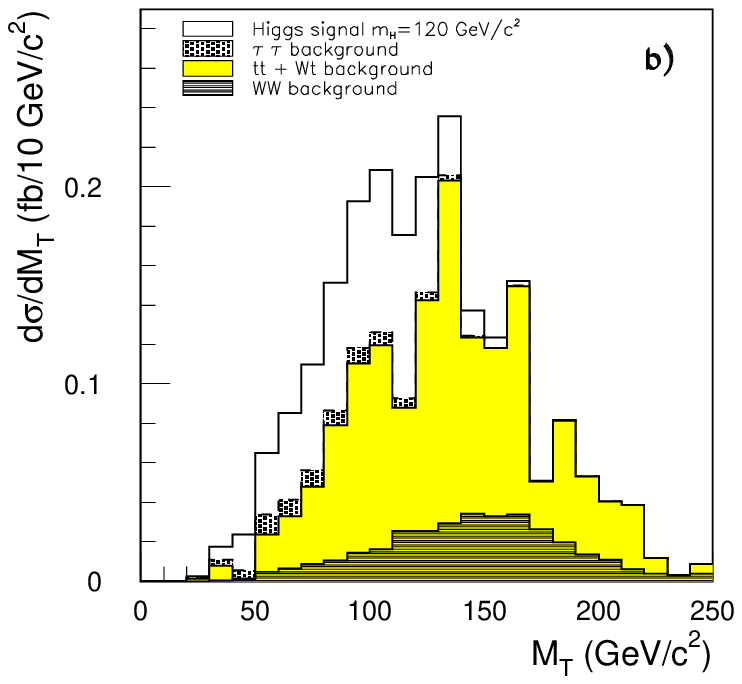,height=7.5cm}}
\end{minipage}
\caption{\small \it 
Distributions of the transverse mass $M_T$ for Higgs boson masses of (a) 160~\Gcs\  
and (b) 120~\Gcs\ in the $e\mu$-channel after all cuts are applied. 
The accepted cross-sections  
$d \sigma / d M_T$ (in fb/10~\Gcs) including all efficiency and acceptance factors 
are shown in both cases.}
\label{f:mt-plot}
\end{center}
\end{figure}

The three final-state topologies have been simulated separately in PYTHIA.
In the generation, a parton-level \PT\ cutoff of 10~\Gc\ has been introduced to 
regulate the divergencies appearing in the tree-level matrix elements. 
If the cross-section contributions of the three processes are   
added, the total \ttbar\ background is estimated to be 1.08 fb, which is 
a factor of 2.1 higher than the value determined with the PYTHIA parton-shower
approach~\cite{jakobs01}. This simple addition may overestimate the 
true \ttbar\ background
since the divergencies in the tree-level calculations are not compensated 
for by \PT\ dependent topological K-factors, which may take values smaller 
than 1 in the low-\PT\ region. Therefore, this estimate is used in the
evaluation of the signal significance as a conservative estimate of the 
$\ttbar \ + \ jet $ background. 
The distribution of the transverse mass $M_T$, assuming this background estimate, 
is shown in Fig.~\ref{f:mt-plot} for Higgs boson signals of 160~\Gcs\ and 120~\Gcs.

Given the major uncertainties in predicting the absolute level of the 
\ttbar\ background, it is important to determine this background directly in
the experiment. 
As mentioned already above, only a transverse mass peak can be reconstructed.
After all cuts are applied, most background events lie 
in the same region of transverse mass as the signal. It is expected that at least 
the shape of the dominant \ttbar\ background can be determined from \ttbar\ events 
observed at the LHC. In \ttbar\ events one may require only one leptonic decay 
and reconstruct the second top decaying into three jets.
After requiring the forward jet tag criteria in addition, 
and correcting for differences in reconstruction efficiencies, this should 
provide an absolute prediction of the background rate with 
small uncertainties.

\begin{figure}[hbtn]
\begin{center}
\begin{minipage}{7.7cm}
\mbox{\epsfig{file=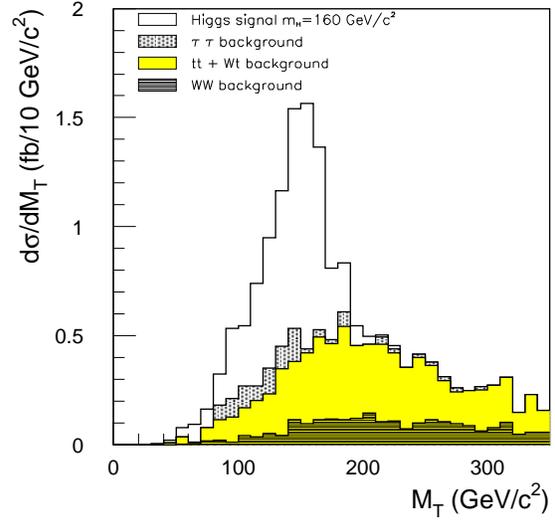,height=7.5cm}}
\end{minipage}
\caption{\small \it 
Distribution of the transverse mass $M_T$ for a Higgs boson mass of 160~\Gcs\ 
in the $e\mu$-channel
above the sum of the various backgrounds after all cuts 
except the lepton cuts (see Table~\ref{t:wwcuts}) are applied.  
The accepted cross-section  
$d \sigma / d M_T$ (in fb/10~\Gcs) including all efficiency and 
acceptance factors is shown.}
\label{f:mt-nolep}
\end{center}
\end{figure}

In addition, the selection cuts can be varied. An interesting approach is, 
for example, to apply all cuts discussed above except the lepton 
cuts, as defined in Table~\ref{t:wwcuts}. The distribution of the 
reconstructed transverse mass 
$M_T$ is shown for a Higgs boson signal with $m_H = 160$~\Gcs\ 
above the backgrounds in Fig.~\ref{f:mt-nolep} 
without 
applying the lepton cuts. Even if no lepton cuts are applied, a 
clear signal can be seen above a background which now extends to higher 
$M_T$ values. The background in the high-$M_T$ region can be used to 
perform a normalization outside of the signal region and to 
predict the background below the signal peak, if the shape of the 
distribution is taken from a Monte Carlo prediction. 
Already for an integrated luminosity of 10 \fbs, a statistical uncertainty 
of about 10\% for the predicted background can be reached. In addition, as 
discussed in the following subsection, the distribution of the azimuthal difference 
$\Delta \phi$ between the two leptons can be used to extract a background 
normalization for events below the signal peak. Given these additional 
possibilities, it is conservatively assumed in the following that the total 
background rate can be determined with an uncertainty of $\pm 10\%$.

\subsubsection{Consistency with a spin-0 resonance in \hwws\ decays}
\begin{figure}[hbtn]
\begin{center}
\begin{minipage}{7.7cm}
\mbox{\epsfig{file=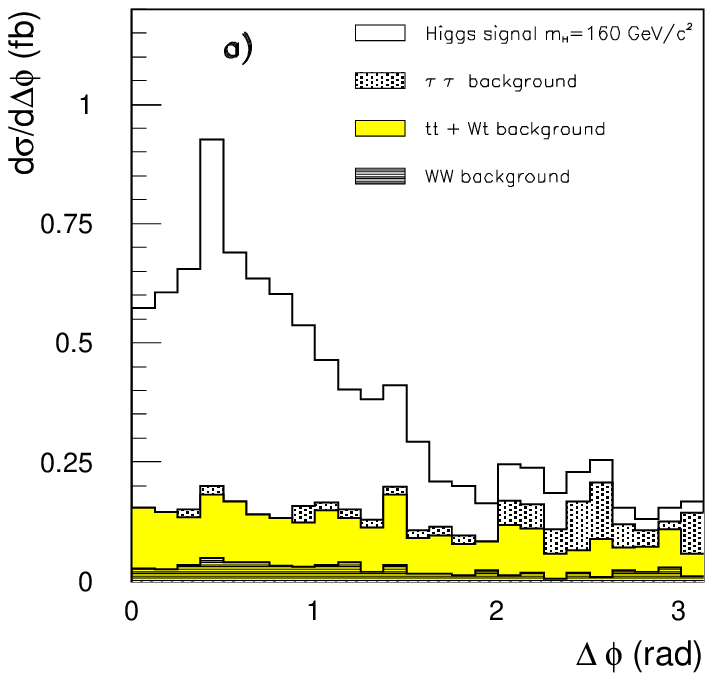,height=7.5cm}}
\end{minipage}
\begin{minipage}{7.7cm}
\mbox{\epsfig{file=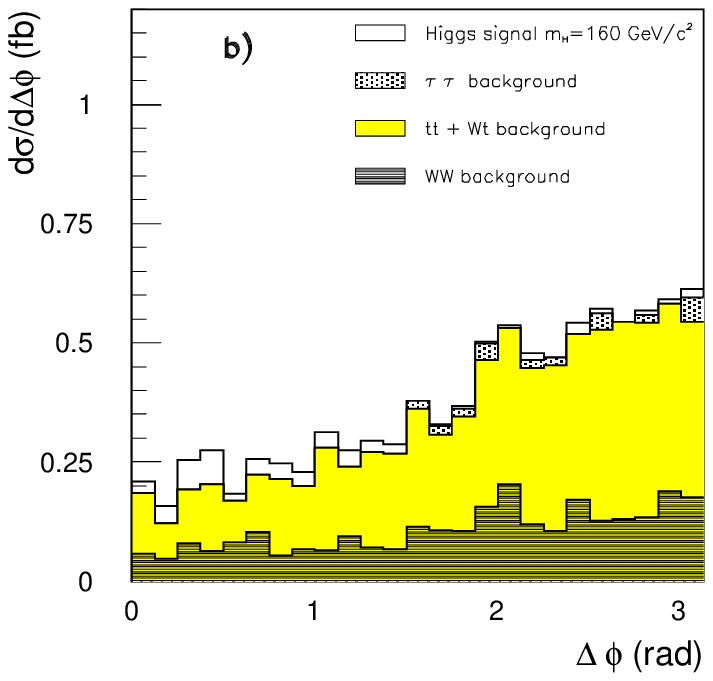,height=7.5cm}}
\end{minipage}
\caption{\small \it 
Distributions of the azimuthal opening angle $\Delta \phi$ 
between the two leptons for (a) events in the signal region ($M_T < 175$~\Gcs) 
and (b) events outside the signal region ($M_T > 175$\Gcs). 
The accepted cross-sections  
$d \sigma / d M_T$ (in fb/10~\Gcs) including all efficiency and 
acceptance factors are shown in both cases.}
\label{f:delphi_nolep}
\end{center}
\end{figure}

The relaxed selection criteria can also be used to extract additional 
information on the spin of the Higgs boson in the \hwws\ decay mode. As discussed
above, the di-lepton azimuthal angular separation $\Delta \phi$ is sensitive to the
spin of the Higgs boson. 
The selection obtained without applying the lepton cuts allows an  
unbiased $\Delta \phi$ distribution to be reconstructed.
In Fig.~\ref{f:delphi_nolep},
this distribution is shown for $e\mu$ final states passing all cuts 
except the lepton cuts. For this example, corresponding to a Higgs boson mass
of 160~\Gcs, the events have been separated into two different regions of 
$M_T$, (a) in the so called signal region ($M_T < 175$~\Gcs) and (b) in a control 
region ($M_T > 175$~\Gcs). For events in the signal region, the distribution 
shows the effect of a spin-0 resonance above a flat background. The pronounced 
structure at small $\Delta \phi$ is not present for events in the control  
sample, where the \ttbar\ and $WW$ backgrounds are 
expected to dominate. Therefore, the unbiased $\Delta \phi$ distribution 
in the signal region can be used for both a demonstration of the 
consistency of the signal with a spin-0 hypothesis and for an 
additional background normalization. This normalization can be performed in  
the high $\Delta \phi$ region directly from events below the peak.

\subsection{The $\ell \nu$-jet-jet channel}

In addition to the di-lepton final states, 
a study has been made as to whether the larger branching ratio of
the W-bosons into quark pairs can be used and the process
$ qq \rightarrow qqH \rightarrow qqWW^{(*)} \rightarrow qq \ \ell \nu \ jj$ can be 
identified above the larger backgrounds. This process has already been
established as a discovery channel for a heavy Higgs boson
\cite{atlas-tdr} in the vector boson fusion process, but has so far
not been considered in the intermediate mass region. 

Due to the larger hadronic branching ratio of the W into a $q\bar{q}$-pair, the
cross-section times branching ratio for this channel is about
4.3 times larger than for the di-lepton channel. However, since there is
only one lepton in the final state, $W+jet$ production is a serious
additional background, with a cross-section which is 
more than two orders of magnitude larger than the signal cross-section.
The W+4 jet background has been evaluated 
using both the PYTHIA parton-shower approach and the VECBOS~\cite{vecbos} 
explicit matrix element calculation. The VECBOS prediction has been found to 
be a factor of two larger. Since VECBOS describes well 
the measurements of W+jet production at the TeVatron~\cite{cdfwjets,d0wjets},
it has been used also for the background estimate in this study. 
In order to suppress this large background, very tight cuts on the tag
jets have been applied. The two jets from  the W decay are searched for 
in the central region of the detector, $| \eta | < 2.0$.
The invariant mass of the two jets 
is required to be compatible with the W mass. 
A mass window cut is applied for Higgs boson searches with $m_H >$ 160~\Gcs, where 
two on-shell W's are expected.  For
lower Higgs boson masses, the lower mass cut has been removed since the 
jets may also originate from an off-shell W boson. 
Similar to the di-lepton analysis,
angular correlations resulting from the decay of the scalar Higgs
boson into two vector bosons are exploited. Detailed studies have
shown \cite{pisa-ww} that in this case, cuts on the minimal and maximal 
separation between 
the lepton and the jets from the W decay, $\Delta R_{min} = min {(\Delta R(l,j_1),
\Delta R(l,j_2) )}$ and $\Delta R_{max} = max {(\Delta R(l,j_1),
\Delta R(l,j_2) )}$, lead to an improved signal-to-background ratio. 

The cuts applied in this analysis are: 
\begin{itemize}
\item One isolated lepton with \\
30~\Gc\ $< P_{T} <$ 100~\Gc\ and $| \eta| < 2.5$  and \\
missing transverse momentum $\ptmiss >$ 30~\Gc.

\item Two tag jets with $ \PT > 60$~\Gc, \\  
$ \Delta \eta_{tags} = |\eta_{tag}^{1} - \eta_{tag}^{2} | > 5.0 $ and  \\
$M_{jj} >$ 1200~\Gcs. \\
In addition, the number of jets with $ \PT
> 20$~\Gc\ in the forward region of the detector (2 $< | \eta | <$ 5) 
is required to be smaller than 5.  

\item {Cuts on jets in the central region:} 
two jets in the central region ($| \eta | <$ 2.0) with: \\
30~\Gc\ $< \PT (j_1) <$ 100~\Gc\ and \\ 
25~\Gc\ $< \PT (j_2) <$  75~\Gc\ and \\
65~\Gcs\ $< M(j_1,j_2) <$ 90~\Gcs\  for $m_H \ge 160$~\Gcs\ and \\ 
\hspace*{2.0cm}$ M(j_1,j_2) <$ 90~\Gcs\ for $m_H < 160$~\Gcs. 

\item Jet veto: no additional jets with $\PT > 20$~\Gc\ in the central region 
$| \eta | < 2.0$.

\item Angular cuts: $\Delta R_{min} < 1$ and $\Delta R_{max} < 2$ \\
for $m_H \ge 160$~\Gcs\ and \\
\hspace*{2.5cm} $\Delta R_{min} < 0.8 $ and $\Delta R_{max} < 1.4$ \\ 
for $m_H < 160$~\Gcs. 

\end{itemize}

The acceptance of the cuts for a Higgs boson signal of 160~\Gcs\ and for the 
backgrounds is summarized in Table \ref{t:acc-lnjj}. 

\renewcommand{\baselinestretch}{1.1}\selectfont
\begin{table*}
\begin{center}
\begin{minipage}{0.75\linewidth}
\footnotesize 
\begin{center}
\begin{tabular}{l||r ||r|r|r|r|r }
\hline
\hline
 & signal & \multicolumn{5}{c}{background (fb)} \\
 &VV (fb) & \ttbar\ & W+jet & \multicolumn{2}{c|}{\wwjets} & Total \\
 &  &   &     & EW & QCD & \\
\hline
Lepton acceptance + \ptmiss\      & 40.5  & 3300 & 20000 & 22.3 & 65.3 & 23388   \\
+ Tag jets, incl. mass cut        &  5.8  &  135 &  690  & 11.6 &  0.1 & 837   \\
+ Central Jets                    &  1.6  &  9.0 &  14.9 &  2.5 & $<$0.01& 26.4   \\
+ Jet veto                        &  1.5  &  2.4 &   7.7 &  2.2 & $<$0.01& 12.3  \\
+ Angular Cuts                    &  0.8  &  0.1 &   0.4 &  0.1 & 
$<$0.01& 0.6  \\
\hline
\hline
\end{tabular}
\vspace{0.5cm}
\caption{\small \it Accepted signal (for $m_H =$ 160~\Gcs) and background cross-sections 
 in fb for the $H \rightarrow WW \rightarrow l \nu jj$ channel after the application of 
successive cuts.}
\label{t:acc-lnjj}
\end{center}
\end{minipage}
\end{center}
\end{table*}

As can be seen from these numbers, the final signal rates after all
cuts are expected to be much lower than the corresponding numbers in the 
di-lepton channel, with a much smaller signal-to-background ratio. 
A possible observation of a Higgs boson with a mass around 160~\Gcs\ can be
confirmed in this channel for integrated luminosities around 30~\fbs.
However, there might be large systematic uncertainties on the estimate of
the dominant W+jet background. In addition, 
it must be stressed that very hard cuts on the \PT\ and on the 
invariant mass of the forward tag jets, as well as on the separation 
$\Delta R$ between the lepton and the jets from the W-decay, are 
necessary to extract the signal above the large backgrounds. 
These extreme cuts might also lead to larger systematic uncertainties on the 
background prediction.

\subsection{Discovery potential as a function of mass}

The analyses outlined above has been performed in the full range of Higgs boson 
masses from 110 to 190 \Gcs. As illustrated in Fig.~\ref{f:mt-plot}, Higgs boson 
signal events are mainly reconstructed in the region of transverse mass 
50~\Gcs,$< m_T < m_H + 10$~\Gcs. For Higgs boson masses above the threshold of
two real W bosons ($m_H > 160$ \Gcs) the upper mass bound has to be increased 
to optimize the signal significance. The upper mass bounds 
are given in Table~\ref{t:ww_sig} 
together with the expected numbers of signal and
background events in the corresponding interval of transverse mass 
for all three $WW^{(*)}$ channels considered
for an integrated luminosity
of 10~\fbs\ (respectively 30~\fbs\ for the $\ell \nu \ jet \ jet$-channel).

\begin{table*}
\begin{center}
\begin{minipage}{0.75\linewidth}
\footnotesize 
\begin{center}
\begin{tabular}{l r || c| c | c | c | c | c | c | c | c }
\hline
\hline
$m_H$ & (\Gcs)         & 110 & 120 & 130 & 140 & 150 & 160 & 170 & 180& 190 \\
 & & & & & & & & & & \\
\hline
\hline
Upper $M_T$ bound for & (\Gcs)& 120 & 130 & 140 & 150 & 160 & 175 & 190 & 220& 240 \\  
mass window           &     &     &     &     &     &     &     &    &     \\
\hline
\hline

\multicolumn{2}{c ||}{$ \hwws \rightarrow e \mu + X$}  & & & & & & & & & \\
Signal     & (10 \fbs\ )  & 1.4  & 4.9  & 12.3 &16.3  & 26.2 & 42.5 & 42.7 & 35.6 & 27.8 \\
Background & (10 \fbs\ )  & 5.8  & 7.1  & 9.2  & 8.1  & 9.8  & 12.4 & 13.8 & 16.3 & 17.1 \\
Stat. significance  & (10 \fbs\ )  
                         &  0.5  &  1.5 & 3.2  & 4.2  & 6.0  & 8.1  & 7.8  & 6.3  & 5.0  \\
\hline
\hline
\multicolumn{2}{c||}{$ \hwws \rightarrow ee/\mu\mu + X$} & & & & & & & & &  \\
Signal     & (10 \fbs\ )  & 1.3  & 4.6  &11.7  &16.4  & 27.8 & 40.2 & 44.8 & 36.0 & 25.9 \\
Background & (10 \fbs\ )  & 6.7  & 8.7  &10.1  &10.0  & 12.2 & 14.3 & 15.9 & 18.4 & 19.2 \\
Stat. significance  & (10 \fbs\ )  
                         &  0.4 &  1.3 & 2.9  & 3.9  & 5.8  &  7.4 &  7.7 & 6.1 & 4.5  \\
\hline
\hline
\multicolumn{2}{c||}{$ \hwws \rightarrow l \nu \ jj $} & & & & & & & & & \\
Signal     & (30 \fbs\ )  &  -   &  -   & 4.5  & 7.5  & 10.5 & 24.0 &24.0  &18.0 & 15.0 \\
Background & (30 \fbs\ )  &  -   &  -   & 6.0  & 6.0  &  6.0 & 18.0 &18.0  &18.0 & 18.0 \\
Stat. significance & (30 \fbs\ )  
                   &  -   &  -   & 1.5  & 2.4  &  3.3 &  4.6 &  4.6 & 3.5 & 3.0 \\
\hline
\hline
Combined           & & & & & & & & & & \\
Stat. significance & (10 \fbs\ ) & 0.8  & 2.1  & 4.4  & 5.9 & 8.4 & 11.0 & 11.0 & 8.8 & 6.8 \\
\hline
\hline
\end{tabular}
\vspace{0.5cm}
\caption{\small \it Expected signal and background rates and  signal significance 
for the three $WW^{(*)}$
decay channels as a function of $m_H$ assuming an
integrated luminosity of 10 \fbs\ (30 \fbs\ for the 
$ \hwws \rightarrow l \nu \ jj $ channel). 
The signal significance has been 
computed using Poisson statistics and assuming a systematic uncertainty 
of 10\% on the background. The conservative estimate for the $\ttbar$ 
background has been used.
}\label{t:ww_sig}
\end{center}
\end{minipage}
\end{center}
\end{table*}

\begin{figure*}
\begin{minipage}{1.0\linewidth}
\begin{center}
\mbox{\epsfig{file=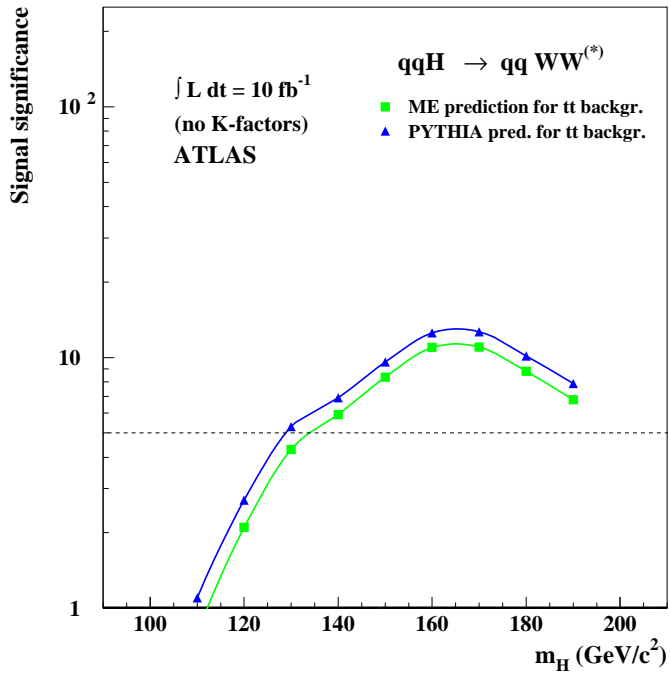,height=10.0cm}}
\caption{\small \it 
ATLAS sensitivity for the discovery of a Standard Model Higgs boson in the 
$qq H \to qq W W $ channel
for an integrated luminosity of 10 \fbs\ . The 
signal significance is plotted for the combination of the three channels considered
and for two estimates of the \ttbar\ background. This background has been computed 
using the PYTHIA parton-shower approach as well as the explicit matrix element
calculation for $\ttbar + 0, 1$, and $2 \ jets$.}
\label{f:ww-signif}
\end{center}
\end{minipage}
\end{figure*}

A signal-to-background ratio larger than $1 $ is found in the di-lepton 
channels for $m_H > 130$ \Gcs. The signal significance, expressed in the 
equivalent number of Gaussian standard deviations, has been calculated 
assuming an integrated luminosity of 10~\fbs, the conservative 
estimate of the \ttbar\ background and a systematic 
uncertainty of 10\% on the level of the background below the signal peak. 
Since the numbers of signal and 
background events are small, Poisson statistics has been used in the evaluation 
of the signal significance. 
With 10 \fbs, a 5\sig\ discovery can be claimed
for $ 145 < m_H < 190$~\Gcs\ in the $e \mu$ channel alone. 
For an integrated luminosity 
of 30 \fbs\ the discovery range increases to 
$125 < m_H < 190$~\Gcs\ if both di-lepton channels are combined. 
Already with an integrated luminosity of 5 \fbs\ only, a mass interval from 
150 - 190~\Gcs\ can be covered. The conservative assuption of the larger 
$\ttbar$ background only slightly affects the discovery potential. 
In Fig.~\ref{f:ww-signif}, the signal significance is shown as a function of the 
Higgs boson mass for both assumptions of the \ttbar\ background and for an 
integrated luminosity of 10~\fbs.

\section{The $ H \rightarrow \tau \tau$ decay mode}

In the following, searches for $H \rightarrow \tau \tau$ decays  
using the  double leptonic decay mode, 
$ qqH  \rightarrow qq \ \tau \tau \rightarrow qq \ l^+ \nu \bar{\nu} \  
l^- \bar{\nu} \nu$ and the lepton-hadron decay mode
$ qqH  \rightarrow qq \ \tau \tau$ $\rightarrow qq$  $l^{\pm} \nu \nu \ had \ \nu$,
are described. 
The analysis strategy follows to a large extent 
the one described above for the $WW$ decay modes. 
The same backgrounds as considered in the 
$WW$ analysis are also relevant here. The $Z + jet$ production, followed by the
$Z \rightarrow \tau \tau$ decay, constitues the principal 
background for $H \rightarrow \tau \tau$ decays at low Higgs boson masses. 
This background has been evaluated using a matrix element based Monte 
Carlo generator. 
The main points of the analyses are briefly 
summarized in the following subsections. For details the reader is referred to 
Refs.~\cite{rachid}, \cite{klute}, \cite{shoji} and \cite{bill}.

\subsection{Di-lepton final states: $H \rightarrow \tau \tau \rightarrow \ell^+ 
\ell^- \ptmiss + X$ }

The following cuts have been applied to select the various di-lepton 
final states: 

\begin{itemize}
\item Two isolated leptons with \\
$ P_{T}(e) > 15 $  \Gc\  and $| \eta_{e}| \le 2.5 $ and \\
$ P_{T}(\mu) > 10$  \Gc\ and $| \eta_{\mu}| \le 2.5$.\\
In addition, a b-jet veto is applied, {\em i.e.}, the event is rejected if 
a b-jet with $\PT > 20$~\Gc\ and  $| \eta | < 2.5$ is identified. A b-jet 
tagging efficiency of 0.6 is assumed~\cite{atlas-tdr}. 

\item Two tag jets with $P_T^1 > 50$ \Gc,  $P_T^2 > 30$ \Gc\  and 
$ \Delta \eta_{tags} = |\eta_{tag}^{1} - \eta_{tag}^{2} | \ge 4.4 $. \\  
In addition, it has been required that the leptons be 
reconstructed within the pseudorapidity gap spanned
by the two tag jets: 
$\eta_{tag}^{min} <  \eta_{l_{1,2}}  < \eta_{tag}^{max}$.

\item Missing transverse momentum: \\
$ \ptmiss > 50$~\Gc. 

\item Invariant mass of the two tag jets: \\
$M_{jj} > 700$~\Gcs.

\item Jet veto: no jets with $\PT > 20$ \Gc\ in $| \eta | < 3.2$. 

\item Azimuthal separation $\Delta \phi_{jj}$ between the tag jets: 
 $\Delta \phi_{jj} < 2.2$.\\
This cut is applied to reduce the electroweak $Zjj$ background, for which 
back-to-back jets are preferred \cite{eboli}.
\item Separation between the two leptons: $\Delta R_{\ell \ell} < 2.6$.

\item Tau reconstruction: \\
$ x_{\tau_{1}}, x_{\tau_{2}} > 0 $, \hspace*{1.0cm} 
$ x_{\tau_{1}}^2 + x_{\tau_{2}}^2 < 1$. 

\item Mass window around the Higgs boson mass: \\
$m_H - 10$~\Gcs\ $< m_{\tau \tau} < m_H + 15$~\Gcs. \\
\end{itemize}

The results of the analysis are summarized in Table \ref{t:tau01}.  The 
accepted cross-sections for the signal with $m_H = 120$ \Gcs\
and the background contributions 
are given after the application of the consecutive cuts for the $e \mu$ channel.
The final numbers are also included for $ee$ and $\mu \mu$ final states.

\renewcommand{\baselinestretch}{1.1}\selectfont
\begin{table*}
\begin{center}
\begin{minipage}{0.75\linewidth}
\footnotesize 
\begin{center}
\begin{tabular}{l||r r||r|r|r|r|r|r}
\hline
\hline
 & \multicolumn{2}{c ||}{signal (fb)} & \multicolumn{6}{c}{background (fb)} \\
 &VV & gg & \ttjets\ & \multicolumn{2}{c|}{\wwjets} & \multicolumn{2}{c|}
{$\gamma^*/Z \ + jets$} & Total \\
 & & & & EW & QCD & EW & QCD & \\
\hline
Lepton acceptance       & 5.55 &       &2014. & 18.2 &669.8 & 11.6 &2150. & 4864.\\
+ Forward Tagging       & 1.31 &       & 42.0 & 9.50 & 0.38 & 2.20 &27.5  &  81.6\\
+ $\ptmiss$             & 0.85 &       & 29.2 & 7.38 & 0.21 & 1.21 &12.4  &  50.4\\
+ Jet mass              & 0.76 &       & 20.9 & 7.36 & 0.11 & 1.17 & 9.38 &  38.9\\
+ Jet veto              & 0.55 &       & 2.70 & 5.74 & 0.05 & 1.11 & 4.56 &  14.2\\
+ Angular cuts          & 0.40 &       & 0.74 & 1.20 & 0.04 & 0.57 & 3.39 &  5.94\\
+ Tau reconstruction    & 0.37 &       & 0.12 & 0.28 & 0.001& 0.49 & 2.84 &  3.73\\
+ Mass window           & 0.27 & 0.01  & 0.03 & 0.02 & 0.0  & 0.04 & 0.15 &  0.24\\
\hline
$H \rightarrow \tau \tau \rightarrow e \mu$
                        & 0.27 & 0.01  & 0.03 & 0.02 & 0.0  & 0.04 & 0.15 &  0.24 \\
\hline
\hline
$H \rightarrow \tau \tau \rightarrow ee $  
                        & 0.13 & 0.01  & 0.01 & 0.01 & 0.0  & 0.02 & 0.07 &  0.11\\
\hline
\hline
$H \rightarrow \tau \tau \rightarrow \mu \mu $ 
                        & 0.14 & 0.01  & 0.01 & 0.01 & 0.0  & 0.02 & 0.07 &  0.11\\
\hline
\hline
\end{tabular}
\vspace{0.5cm}
\caption{\small \it Accepted signal (for $m_H$ = 120~\Gcs) and background 
cross-sections 
in fb for the $H \rightarrow \tau \tau \rightarrow e\mu + X$ channel after the 
application of successive cuts. For the signal,
the contributions via the vector boson fusion and the 
gluon fusion channel are given separately. The final results for  
$ee$ and $\mu \mu$ final states are given in the last rows of the table.}
\label{t:tau01}
\end{center}
\end{minipage}
\end{center}
\end{table*}

A similar rejection pattern as seen in the $\hwws$ analysis can be observed. 
It should be pointed out that the tau reconstruction
with cuts on $x_{\tau_1}$ and $x_{\tau_2}$
leads to a significant suppression of the reducible backgrounds, while the signal 
is kept with high efficiency. After $\tau$ reconstruction the signal-to-background 
ratio is still much smaller than 1. This situation is drastically changed after the 
application of the mass cut around the Higgs boson mass. 
The sidebands can be used for the determination of the absolute 
level of the background. The mass window and therefore the background level depend
on the $\tau \tau$ mass resolution. This resolution has so far only be determined 
with the ATLAS fast detector simulation, where a value of 12~\Gcs\ has been found 
for a Higgs boson with a mass of 120~\Gcs. 
A verification of this mass resolution in a full detector simulation remains to 
be done. However, based on the experience gained in the MSSM decay channels
$A/H \rightarrow \tau \tau $ \cite{donatella}, no big differences between 
fast and full simulations are expected. 

For a Higgs boson with a mass of 120 \Gcs,
the dominant background results from $Z+jet$ production.  
Despite the good signal-to-background ratio found, 
the accepted cross-sections are low and a data sample 
corresponding to 20-30~\fbs\ is needed for a significant observation 
of a Higgs boson in this channel. Therefore, an
observation of the $H \rightarrow \tau \tau$ decay mode at the LHC in the vector 
boson fusion process looks feasible, as suggested 
in Ref.~\cite{zeppenfeld-tau}. 

All numbers given in the upper part of Table~\ref{t:tau01} 
refer to $e \mu$ final states. The final numbers, 
after the application of all cuts, 
for the $ee$ and $\mu \mu$ final states are given in the last two lines of 
the table. The contributions from other sources of Higgs boson 
production have also been evaluated. The contribution from the gluon fusion 
is found to be small, and contributes after all cuts at a level of 
about 4\% of the vector boson fusion cross-section to the Higgs boson signal. 
Also the contributions via $H \to W W^{(*)} \to \ell \ell \ptmiss$ to the signal 
have been evaluated. Due to the strong variation of the $WW^{(*)}$
branching ratio with the Higgs boson mass,  
this contribution is found to be mass dependent. It ranges 
between 0.003~fb for $m_H = 110$~\Gcs\ and 0.03~fb for $m_H = 150$~\Gcs. 
Since for those events the reconstructed $\tau \tau$ mass does not correspond 
to the Higgs boson mass, these contributions have been added to the 
background in the evaluation of teh signal significance.

\begin{figure}[hbtn]
\begin{center}
\mbox{\epsfig{file=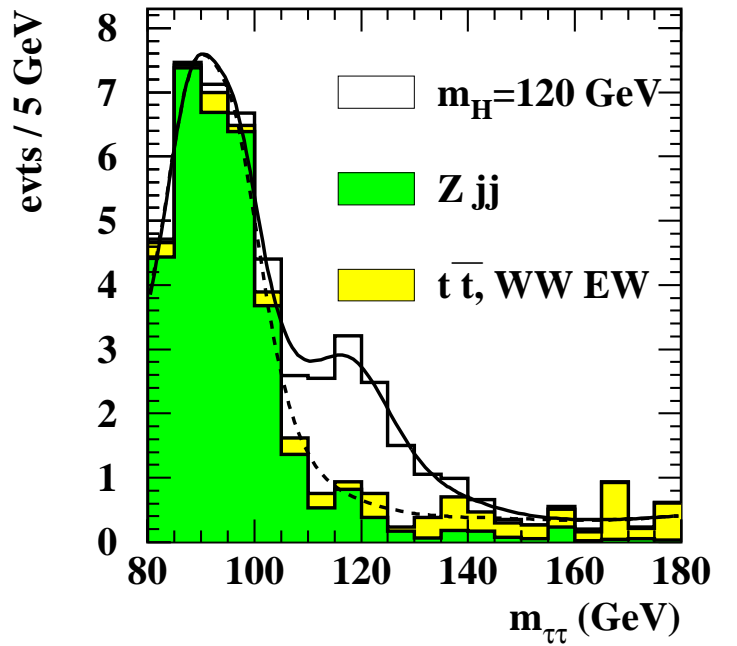,height=8.5cm}}
\caption{\small \it 
The reconstructed $\tau \tau $ invariant mass for a Higgs boson signal 
of 120~\Gcs\
in the $e \mu$ channel  
above all backgrounds after application of all cuts except the mass window cut. 
The number of signal and background events are shown for an integrated 
luminosity of 30 \fbs.}
\label{f:mtautau}
\end{center}
\end{figure}

The distribution of the reconstructed $\tau \tau$ invariant mass for the 
$e \mu$ final state is shown 
in Fig.~\ref{f:mtautau}  
for a Higgs boson signal of 120~\Gcs\ above the background for an 
integrated luminosity of 30~\fbs. 

\subsection{The lepton-hadron decay mode:
$H  \rightarrow \  \tau \tau \rightarrow  \ \ell \nu \nu \ had \ \nu$}

The lepton-hadron final states are 
selected by applying the following cuts: 

\begin{itemize}
\item One isolated lepton ($e$ or $\mu$) with \\
$ P_{T}(e) > 25 $  \Gc\  and $| \eta_{e}| \le 2.5 $ or \\
$ P_{T}(\mu) > 20$ \Gc\  and $| \eta_{\mu}| \le 2.5$.

\item One identified hadronic $\tau$ cluster in the calorimeter
with $\PT > 40$ \Gc\ passing the standard ATLAS tau selection 
criteria~\cite{atlas-tdr,donatella}.
In the present analysis, a selection with an 
efficiency of 50\% for hadronic tau decays has been used. 

\item Two tag jets with $P_T^1 > 40$ \Gc,  $P_T^2 > 20$ \Gc\  and 
$ \Delta \eta_{tags} = |\eta_{tag}^{1} - \eta_{tag}^{2} | \ge 4.4 $. \\  
In addition, it has been required that the visible tau decay products be 
reconstructed within the pseudorapidity gap spanned
by the two tagging jets: 
$\eta_{tag}^{min} <  \eta_{\ell,h}  < \eta_{tag}^{max}$.

\item Tau reconstruction: \\
$ 0 < x_{\tau_{\ell}} < 0.75$ and $ 0 < x_{\tau_{h}} < 1 $, \\
where $x_{\tau_{\ell}}$ and $x_{\tau_{h}}$ are the corresponding momentum 
fractions carried by the lepton or the hadronic tau system respectively. 

\item Transverse mass cut: in order to suppress the \ttbar\ background, where 
the lepton and the missing transverse momentum originate from the decay of a 
$W$, a cut on 
the transverse mass \\
$m_T (\ell \nu) = \sqrt{2 P_T(\ell) \ptmiss \cdot ( 1 - cos \Delta \phi )}$ 
is applied: \\
$m_T (\ell \nu) < 30$~\Gcs. 

\item Missing transverse momentum: \\
$ \ptmiss > 30$~\Gc. 

\item Invariant mass of the two tag jets: \\
$M_{jj} > 700$~\Gcs. 

\item Jet veto: no jets with $\PT > 20$~\Gc\ in the pseudorapidity range 
defined by the two tag jets $\eta_{tag}^{min} < \eta_j^{veto} < 
\eta_{tag}^{max}$.

\item Mass window around the Higgs boson mass: \\
$m_H - 10$ \Gcs\ $< m_{\tau \tau} < m_H + 15$ \Gcs. 
\end{itemize}

The results of the analysis are summarized in Table \ref{t:tau02}, where the 
accepted cross-sections for the signal with $m_H = 120$~\Gcs\ 
and the background contributions are given after the application of the 
consecutive cuts. 

\renewcommand{\baselinestretch}{1.1}\selectfont
\begin{table*}
\begin{center}
\begin{minipage}{0.75\linewidth}
\footnotesize 
\begin{center}
\begin{tabular}{l||r r||r|r|r|r|r}
\hline
\hline
 & \multicolumn{2}{c||}{signal (fb)} & \multicolumn{4}{c}{background (fb)} \\
 &VV & gg & \ttbar\ & \multicolumn{2}{c|}{$\gamma^*/Z \ + jets$}& W+jet&Total \\
& & & & QCD & EW &  \\
\hline
Lepton acceptance       &13.7  & 50.3  & 1.6$\cdot 10^4$&6925 & 22.0 &
3.4$\cdot 10^4$   & 5.7$\cdot 10^4$  \\
+ Identified had. $\tau$& 6.18 & 22.7  & 4274. & 1842 & 8.03 & 3200. & 9462   \\
+ Forward Tagging       & 1.97 & 0.18  & 29.7  & 23.6 & 1.72 & 30.0  &  85.0  \\
+ Tau reconstruction    & 1.27 & 0.11  & 6.06  & 13.8 & 1.09 & 5.9   &  26.9  \\
+ Transverse mass       & 1.02 & 0.07  & 1.74  & 11.9 & 0.92 & 0.63  &  15.2  \\
+ $\ptmiss$             & 0.81 & 0.05  & 1.38  & 8.31 & 0.71 & 0.58  &  11.0  \\
+ Jet mass              & 0.71 & 0.03  & 1.01  & 6.63 & 0.69 & 0.37  &  8.70  \\
+ Jet veto              & 0.63 & 0.02  & 0.14  & 4.24 & 0.66 & 0.21  &  5.25  \\
+ Mass window           & 0.52 & 0.01  & 0.01  & 0.19 & 0.06 & $<$0.01 & 0.27  \\
\hline
\hline
\end{tabular}
\vspace{0.5cm}
\caption{\small \it Accepted signal (for $m_H =$ 120~\Gcs) and background cross-sections
 in fb for the $H \rightarrow \tau \tau \rightarrow \ \ell \nu \nu \ had \ \nu$
channel after the 
application of successive cuts. For the signal, 
the contributions via the vector boson fusion and the 
gluon fusion channel are given separately.}
\label{t:tau02}
\end{center}
\end{minipage}
\end{center}
\end{table*}

A similar rejection pattern as for the di-lepton decay mode can be observed.
Also in this channel, the $Z+jet$ production constitutes the largest background
and a signal-to-background ratio larger than 1 can only be 
achieved after the final mass window cut has been applied. Also for this channel 
the mass resolution has been determined with the ATLAS fast 
detector simulation and has been found to be 11~\Gcs\ for a 
Higgs boson with a mass of 120~\Gcs. 

\begin{figure}[hbtn]
\begin{center}
\mbox{\epsfig{file=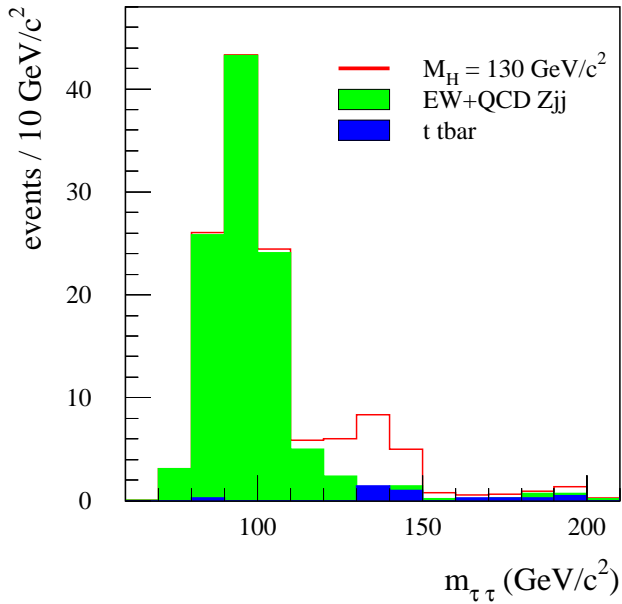,height=8.5cm}}
\caption{\small \it 
The reconstructed $\tau \tau $ invariant mass for a Higgs boson signal of 
130~\Gcs\ in the $\ell - had$ channel  
above all backgrounds after application of all cuts except the mass window cut. 
The number of signal and background events are shown for an integrated 
luminosity of 30 \fbs.}
\label{f:mtau-lhad}
\end{center}
\end{figure}

The distribution of the reconstructed $\tau \tau$ invariant mass for the 
$\ell-had$ final state is shown 
in Fig.~\ref{f:mtau-lhad}  
for a Higgs boson mass of 130~\Gcs\ above the background for an integrated luminosity
of 30~\fbs . 

\subsection{Systematic uncertainties for the $H \rightarrow \tau \tau $ channels}

Since a $\tau \tau$ mass peak can be reconstructed, the absolute level of the 
background can be estimated from sidebands in the distribution. 
This is assumed to work well
in the mass region $m_H > 125$ \Gcs, since there the Higgs mass peak is well 
separated from the $Z \rightarrow \tau \tau $ background peak. For masses in the 
range between 110 and 125~\Gcs, the level still can be determined from a 
normalization to the $Z \rightarrow \tau \tau $ background peak. However, in this 
case uncertainties on the shape of the sum of the  
$Z \rightarrow \tau \tau $ resonance and the remaining \ttbar\ background are 
important. Again, it is assumed that the background can be measured from real 
data after variations of cuts have been performed.
In the evaluation of the signal significance, it is assumed 
that 
after normalization of the $Z \rightarrow \tau \tau $ background at the 
Z-peak, the total background in the mass window is known with an 
uncertainty of $\pm$ 10 \%. 

A comparison between the predictions from the PYTHIA parton-shower Monte 
Carlo and the explicit matrix element calculations has been performed also 
for the dominant $Z + jet$ background. In PYTHIA, only diagrams with one hard jet 
($q\bar{q} \to g \tau \tau$ and $qg \to q \tau \tau$) are available. A second jet, 
if present, arises from initial- or final-state radiation and is softer than jets 
obtained from a matrix element evaluation of the QCD Zjj process. 
Studies have shown~\cite{rachid,shoji} that QCD Zjj events produced with matrix
element calculations yield a background contribution, which is about a factor of 
two larger than the PYTHIA estimate. Therefore, the more conservative 
estimate has been used in the evaluation of the signal significance.

\begin{table}
\footnotesize 
\begin{center}
\begin{tabular}{l r || c| c | c | c | c }
\hline
\hline
$m_H$ & (GeV) & 110 & 120 & 130 & 140 & 150 \\
\hline
\hline
\multicolumn{2}{l ||}{$H \rightarrow \tau \tau \rightarrow e \mu \ptmiss $}  & & & & &  \\
Signal              &  & 9.7  & 8.4  & 6.3  & 3.8  & 1.8  \\
Background          &  &17.3  & 7.1  & 4.1  & 3.0  & 2.6  \\
Stat. significance  &  & 1.9  & 2.5  & 2.3  & 1.7  & 0.7  \\
\hline
\hline
\multicolumn{2}{l ||}{$H \rightarrow \tau \tau \rightarrow ee / \mu\mu \ptmiss$}  & & & & &  \\
Signal              &  & 9.7  & 8.3  & 6.3  & 3.8  & 1.8 \\
Background          &  & 16.2 & 6.6  & 4.5  & 3.5  & 2.6 \\
Stat. significance  &  & 1.9  & 2.6  & 2.3  & 1.5  & 0.7 \\
\hline
\hline
\multicolumn{2}{l||}{$ H \rightarrow \tau \tau \rightarrow \ell \ had \ \ptmiss$} & & & & & \\
Signal              &  & 16.8 & 15.6 & 11.8 & 8.9 & 3.8   \\
Background          &  & 31.9 & 7.7  &  3.6 & 2.5 & 2.5   \\
Stat. significance  &  & 2.4  & 4.2  & 4.4  & 3.9 & 1.7   \\
\hline
\hline
\multicolumn{2}{l||}{ combined } & & & & & \\
Stat. significance  &  & 3.7  & 5.7  & 5.7  & 4.8 & 2.4   \\
\hline
\hline
\end{tabular}
\vspace{0.5cm}
\caption{\small \it Expected signal and background rates and statistical significance 
for the three $\tau \tau$
decay channels as a function of $m_H$ assuming an
integrated luminosity of 30~\fbs.}\label{t:tt_sig}
\end{center}
\end{table}

\subsection{Discovery potential as a function of mass}

The analyses outlined above has been performed for Higgs boson 
masses in the range from 110 to 150~\Gcs. The expected numbers of signal and
background events and the statistical significance for a Higgs boson discovery 
expressed in terms of Gaussian standard deviations are given 
in Table~\ref{t:tt_sig} for an integrated luminosity
of 30 \fbs\ for the three $\tau$ decay channels considered, as well as for the 
combination of all channels. 
As described above, an uncertainty of $\pm 10\%$ on the background is assumed. 
After combination of all channels, an 
observation of a Higgs boson in the $\tau \tau$ decay mode looks feasible 
over the mass range from 115 to 140~\Gcs\ with an integrated luminosity of 
30~\fbs. 

\begin{figure*}
\begin{minipage}{1.0\linewidth}
\begin{center}
\mbox{\epsfig{file=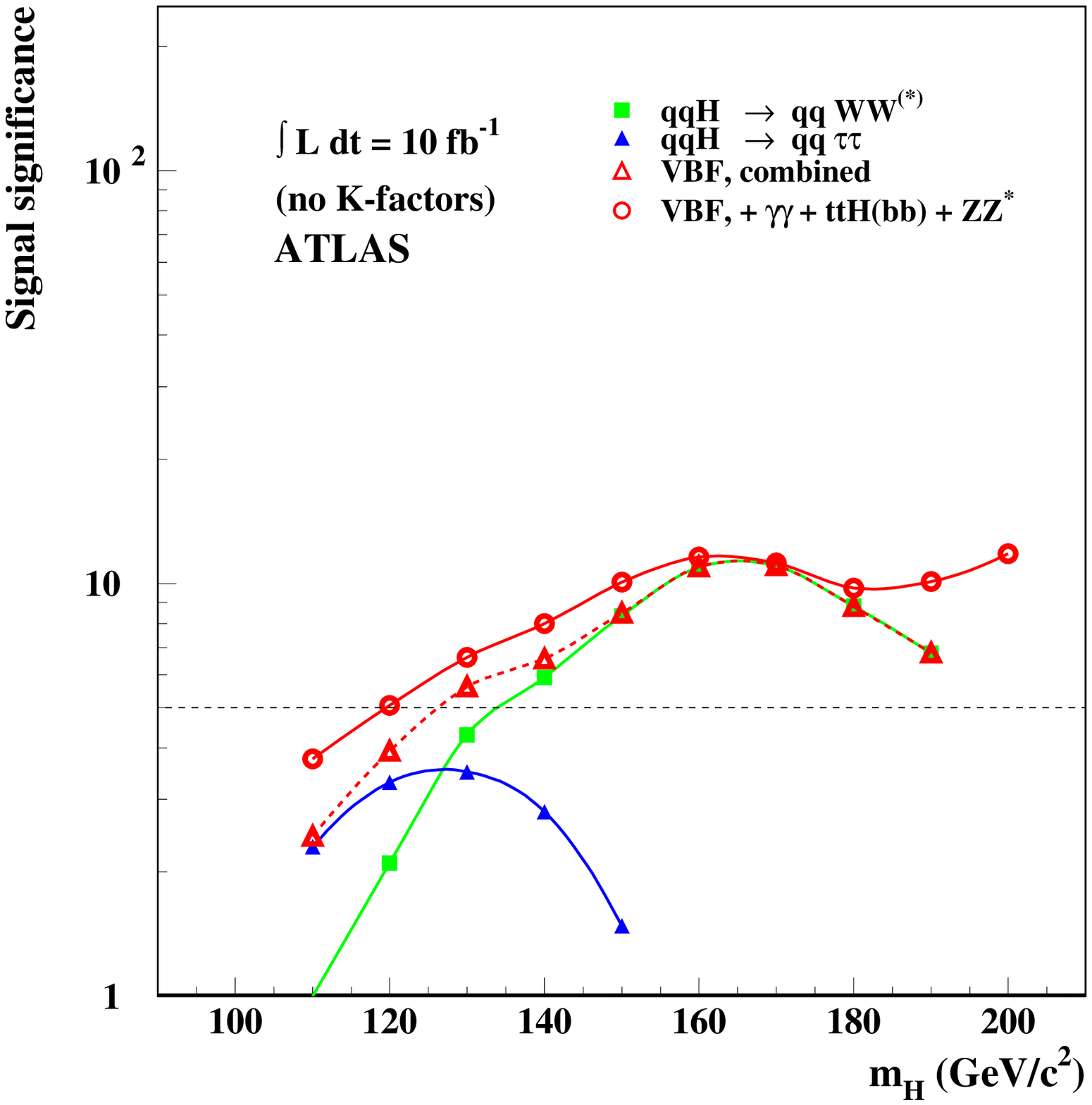,height=10.0cm}}
\mbox{\epsfig{file=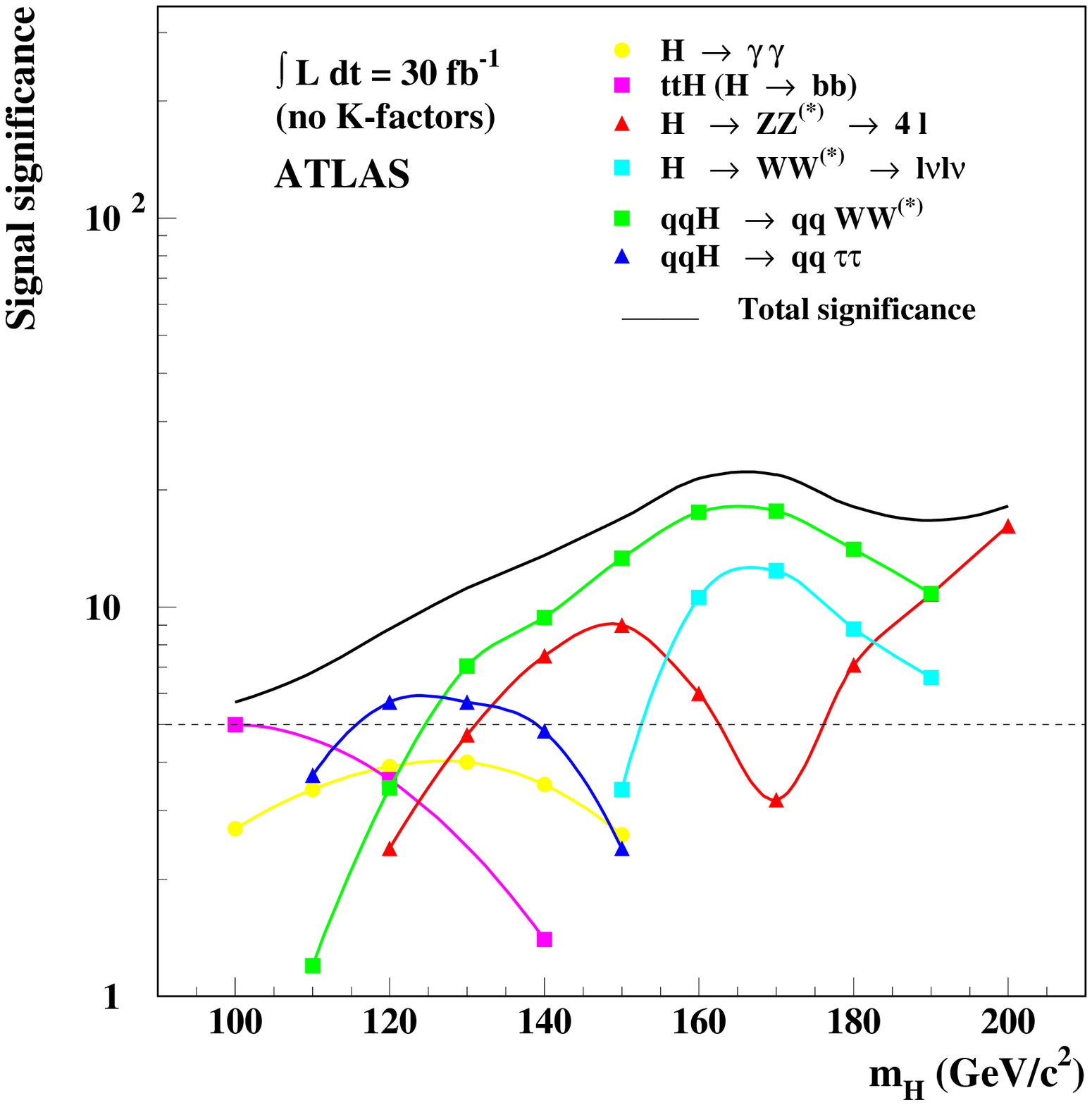,height=10.0cm}}
\caption{\small \it 
ATLAS sensitivity for the discovery of a Standard Model Higgs boson for integrated 
luminosities of 10 and 30 \fbs\ . The 
signal significances are plotted for individual channels, as well as for the
combination of channels. A systematic uncertainty of $\pm$ 10\% on the background
has been included for the vector boson fusion channels.}
\label{f:signif01}
\end{center}
\end{minipage}
\end{figure*}

The number of signal and background events in the selected mass window is affected
by the $\tau \tau$ mass resolution, {\em e.g.}, if the mass resolution would be degraded 
from 12 to 15~\Gcs\, the number of signal events for a Higgs boson with a mass of 
120~\Gcs\ in the $e\mu$-channel would be reduced from 8.4 to 6.8, whereas the number 
of background events would increase from 7.1 to 10.4.

\section{The ATLAS Higgs boson discovery potential}
The ATLAS Higgs boson discovery potential in the mass range 110 - 190~\Gcs\ 
including the vector boson fusion 
channels discussed here is shown in Fig.~\ref{f:signif01} for 
integrated luminosities of 
10 and 30~\fbs. The vector boson fusion channels provide a large discovery 
potential even for small integrated luminosities. Combining the two vector 
boson fusion channels, a Standard Model Higgs boson can be discovered with a 
significance above 5 $\sigma$ in the mass range 135 to 190~\Gcs\ assuming an 
integrated luminosity of 10~\fbs\ and a systematic uncertainty of 10\% 
on the background. If the 
vector boson fusion channels are combined with the standard Higgs boson 
discovery channels~\cite{atlas-tdr} $H \rightarrow \gamma \gamma $, 
$H \rightarrow Z Z^{(*)} \to 4 \ell$, and 
$\ttbar H$ with $ H \rightarrow b \bar{b}$, the 5 $\sigma$ discovery range can 
be extended down to $\sim$120~\Gcs. 

For an integrated luminosity of 30~\fbs, the full mass range can be covered 
by ATLAS with a significance exceeding 5$\sigma$. Over the full mass range 
several channels will be available for a Higgs boson discovery. The various 
discovery channels are complementary both from physics and detector aspects. 
The three different channels test three different production mechanisms,
the gluon-gluon fusion via the $\gamgam$ channnel, the vector boson fusion
via the channels discussed here and the associated $\ttbar H$ 
production via the $\ttbar H$ with $ H \rightarrow b \bar{b}$ mode. 
This complementarity also provides sensitivity to non-standard Higgs models, 
such as fermiophobic models \cite{fermiophobic}. 

Also different detector components are essential for the different channels. 
The $H \rightarrow \gamgam$ decays require excellent electromagnetic 
calorimetry. In the identification of the vector boson fusion the measurement
of jets, in particular the reconstruction of the forward tag jets is essential. 
The Higgs detection in \bbbar\ decays via the associated $\ttbar H$ production   
relies to a large extent on the excellent b-tagging performance of the ATLAS
Inner Detector.

\section{Conclusions}

The discovery potential for the Standard Model Higgs boson in the
mass range below 190~\Gcs\ has been studied using the
vector boson fusion process. It has been demonstrated that the 
ATLAS experiment has a large discovery potential in the 
$ \hwws \rightarrow \ell^+ \ell^- \ptmiss $ channel. The additional signatures
of tag jets in the forward region and of a low jet activity in the
central region of the detector allow for a significant background 
rejection, such that a better signal-to-background ratio than in 
the inclusive $H \rightarrow WW^{(*)}$ channel, which is dominated by the 
gluon-gluon fusion process, is obtained. As a consequence, 
the signal sensitivity is less affected
by systematic uncertainties in the predictions of the background 
levels. The present 
study shows that the ATLAS
experiment at the LHC would be sensitive to a Standard Model Higgs
boson in this decay channel in the mass range between $\sim$135 and 190~\Gcs\ 
with data corresponding to an integrated luminosity of only 10~\fbs.

In addition, it has been shown that in the mass region 
$m_H < 140$~\Gcs, the ATLAS experiment is also sensitive to 
the $\tau \tau$ decay mode of the Standard Model Higgs 
boson in the vector boson fusion channel. 
A discovery in this
final state would require an integrated luminosity of about 
30~$\fbs$.
The detection of the $\tau$ decay mode is 
particularly important for a measurement of the Higgs boson couplings
to fermions. 

The present study confirms 
that the search for vector boson fusion at the LHC has a large discovery
potential over the full range from the lower limit set by the LEP 
experiments up to  $2 m_Z$, where the high sensitivity 
$H \rightarrow  Z Z \ \rightarrow 4 \ \ell$ channel takes over.

\subsection*{Acknowledgements}
This work has been performed within the ATLAS Collaboration, and we thank 
collaboration members for helpful discussions. We have made use of the 
physics analysis framework and tools which are the result of 
collaboration wide efforts. We would like to thank D.~Zeppenfeld, D.~Rainwater, 
T.~Plehn and F.~Gianotti for very useful discussions.


\begin{thebibliography}{99}

\bibitem{atlas-tdr} ATLAS Collaboration, {\em Detector and Physics Performance Technical 
Design Report}, CERN/LHCC/99-14 (1999).

\bibitem{cms-tp} CMS Collaboration, {\em CMS Technical proposal}, CERN/LHCC 94-38, CERN (1994).

\bibitem{LEP-limit} LEP Collaborations, CERN-EP/2001-055 (2001), hep-ex/0107029.

\bibitem{zeppenfeld} D.L.Rainwater and D.Zeppenfeld, 
J. High Energy Phys. 12 (1997) 5, hep-ph/9712271.

\bibitem{zeppenfeld-ww} D.L.Rainwater and D.Zeppenfeld, Phys. Rev. D60 (1999) 113004,
hep-ph/9906218.   

\bibitem{zeppenfeld-tau} D.L.Rainwater, D.Zeppenfeld, K.Hagiwara, 
Phys. Rev. D59 (1999) 14037, hep-ph/9808468;
T.Plehn, D.L.Rainwater and D.Zeppenfeld, Phys. Rev. D61 (2000) 093005.

\bibitem{spira-hqq} M.Spira, VV2H programme, home.cern.ch/m\-/mspira\-/www/proglist.html.

\bibitem{fusion-nlo} T.Han, G.Valencia and S.Willenbrock, Phys. Rev. Lett. 69 
(1992) 3274.

\bibitem{hdecay} M.Spira et al., Comp. Phys. Comm. 108 (1998). 

\bibitem{cteq-sf} H.L.Lai et al., Eur. Phys. J. C12 (2000) 375. 

\bibitem{costanzo_1} V.Cavasinni, D.Costanzo, 
{\em Search for $WH \to WWW \to \ell \nu \ell \nu \ jet \ jet $},
ATLAS internal note ATL-PHYS-2000-013 (2000); \\
D.Cavalli, {\em Combined analysis of $A \to \tau \tau$ events from direct and 
associated $bbA$ production}, ATLAS internal note ATL-PHYS-2000-005 (2000).

\bibitem{pythia} T.Sj\"ostrand, Comp. Phys. Comm. 82 (1994). 

\bibitem{tauola} S.Jadach, Z.Was, CERN-TH-6793 (1993).

\bibitem{rachid} G.Azuelos and R.Mazini, 
{\em Searching for $H \to \tau \tau \to \ell \nu \nu \ + \ h X$ by vector 
boson fusion in ATLAS}, ATLAS internal note ATL-PHYS-2003-004 (2003).

\bibitem{onetop} ONETOP Monte Carlo programme,  
         http://www.pa.msu.edu/brock/atlas-1top/EW-top-programs.html.

\bibitem{comphep} A.Pukhov et al., hep-ph/9908288. 

\bibitem{atlfast} E.Richter-Was, D.Froidevaux, L.Poggioli,
{\em ATLFAST 2.0, a fast simulation package for ATLAS}, 
ATLAS internal note ATL-PHYS-98-131 (1998).

\bibitem{atlas-trigger} ATLAS Collaboration, CERN/LHCC/2000-17 (2000). 

\bibitem{donatella} D.Cavalli and S.Resconi, 
{\em Tau-jet separation in the ATLAS detector}, 
ATLAS internal note ATL-PHYS-98-118 (1998).

\bibitem{jakobs01} C.Buttar, R.Harper, and K.Jakobs,
{\em Weak boson fusion \hwwsll\ as a search mode for an intermediate mass Standard 
Model Higgs boson at ATLAS},  
ATLAS internal note ATL-PHYS-2002-033 (2002).

\bibitem{pisa-tagging} V.Cavasinni, D.Costanzo, I.Vivarelli,
{\em Forward tagging and jet-veto studies for Higgs events produced via vector boson 
fusion}, 
ATLAS internal note ATL-PHYS-2002-008 (2002). 

\bibitem{97ditt} M.Dittmar and H.Dreiner, Phys. Rev. D55 (1997) 167.

\bibitem{mellado} K.Cranmer, P.McNamara, B.Mellado, W.Quayle, Sau Lan Wu, 
{\em Search for Higgs bosons decay $\hwws \to \ell^+ \ell^- \ptmiss$ for 
115 $< M_H <$ 130 GeV using vector boson fusion},
ATLAS internal note ATL-PHYS-2003-002 (2003). 

\bibitem{jakobs-ww} K.Jakobs and Th.Trefzger, 
{\em Standard Model Higgs boson search for \hwwsll\ with a mass between 150 and 
190 GeV}, ATLAS internal note ATL-PHYS-2000-015 (2000).

\bibitem{pisa-ww} V.Cavasinni, D.Costanzo, E.Mazzoni, I.Vivarelli, 
{\em Search for an intermediate mass Higgs boson produced via vector boson fusion in
the channel $\hwws \to \ell \nu \ jet \ jet$},
ATLAS internal note ATL-PHYS-2002-010 (2002).

\bibitem{vecbos} F.A.Berends, W.T.Giele, K.Kuijf, B.Tausk, Fermilab-Pub-90/213-T.

\bibitem{cdfwjets} CDF Collaboration, F.Abe et al., Phys. Rev. Lett. 79 (1997) 4760.

\bibitem{d0wjets} D0 Collaboration, B.Abbott et al., Phys. Lett. B513 (2001) 292.

\bibitem{klute} M.Klute, 
{\em A study of the weak boson fusion with $H \to \tau \tau$ and $\tau \to e(\mu)$},
ATLAS internal note ATL-PHYS-2002-018 (2002).

\bibitem{shoji} T.Takemoto, S.Asai, J.Kanzaki, R.Tanaka {\em Study of 
$H \to \tau \tau \to \ell \ had \ + x$ via vector boson fusion in ATLAS}, 
ATLAS internal note ATL-COM-PHYS-2003-007 (2003).

\bibitem{bill} K.Cranmer, B.Mellado, W.Quayle, Sau Lan Wu, 
{\em An update of the VBF $H \to \tau \tau \to \ell \ell$ cut analysis}, 
ATLAS internal note ATL-COM-PHYS-2003-002 (2003). 

\bibitem{eboli} O.J.Eboli and D.Zeppenfeld, Phys. Lett. B495 (2000) 147. 

\bibitem{fermiophobic} see for example: L.Br\"ucher, R.Santos, hep-ph/9907434, 
and references therein. 

\end{thebibliography}
\end{document}